\documentclass[amsmath,aps,prd,reprint,showpacs,superscriptaddress]{revtex4-1}

\usepackage{amssymb}
\usepackage{braket}
\usepackage{color}
\usepackage{mathtools}
\usepackage{subcaption}
\usepackage{tensor}

\newcommand{\abs}[1]{\left|#1\right|}
\newcommand{\vect}[1]{\boldsymbol{#1}}
\newcommand{\diff}[2][]{\mathrm{d}^{#1} #2}
\newcommand{\pd}[2][]{\operatorname{\partial}_{#2}^{#1}}

\newcommand{\dd}[2][]{\frac{\operatorname{d}^{#1}}{\operatorname{d}#2^{#1}}}
\newcommand{\dadj}[1]{\overline{#1}}
\newcommand{\co}[1]{#1^{*}}

\DeclareMathOperator{\sech}{sech}
\DeclareMathOperator{\artanh}{artanh}
\DeclareMathOperator{\arcosh}{arcosh}

\begin{document}

\title{WKB approach to pair creation in spacetime-dependent fields -- the case
  of a spacetime-dependent mass}

\author{Johannes Oertel}
\email{johannes.oertel@uni-due.de}
\affiliation{Fakult\"at f\"ur Physik, Universit\"at Duisburg-Essen,
  Lotharstra{\ss}e 1, 47057 Duisburg, Germany,}

\author{Ralf Sch\"utzhold}
\affiliation{Fakult\"at f\"ur Physik, Universit\"at Duisburg-Essen,
  Lotharstra{\ss}e 1, 47057 Duisburg, Germany,}
\affiliation{Helmholtz-Zentrum Dresden-Rossendorf,
  Bautzner Landstra{\ss}e 400, 01328 Dresden, Germany,}
\affiliation{Institut f\"ur Theoretische Physik,
  Technische Universit\"at Dresden, 01062 Dresden, Germany.}

\date{\today}

\begin{abstract}
Besides tunneling in static potential landscapes, for example, the WKB approach
is a powerful non-perturbative approximation tool to study particle creation due
to time-dependent background fields, such as cosmological particle production or
the Sauter-Schwinger effect, i.e., electron-positron pair creation in a strong
electric field.
However, our understanding of particle creation processes in background fields
depending on both space and time is rather incomplete.
In order to venture first steps into this direction, we propose a generalization
of the WKB method to truly spacetime-dependent fields and apply it to the case
of a spacetime-dependent mass.
\end{abstract}

\maketitle

\section{Introduction}

Particle creation out of the vacuum due to extreme external influences is an
intriguing effect and a fundamental prediction of quantum field theory.
In the following, we shall focus on electron-positron pair production in quantum
electrodynamics.
There are several possibilities for pair-producing external fields.
For example, in the Sauter-Schwinger
effect~\cite{Sauter1931,*Sauter1932,Heisenberg1936,Schwinger1951}, particles are
created due to a strong electric field.
This is even possible for slowly varying electric fields (as long as they are
strong enough).
Note that this process is different from pair creation in the (perturbative)
multiphoton regime which requires sufficiently fast varying electromagnetic
fields, see, e.g.,~\cite{Burke1997}.
As another example, cosmological pair
production~\cite{Schrodinger1939,Parker1968} occurs in an expanding or
contracting universe.

So far, electron-positron pair production has been verified experimentally only
in the perturbative (multiphoton) regime~\cite{Burke1997}.
Non-perturbative pair production due to an external field is far more difficult
to observe in nature and also not nearly as well understood on the theoretical
side.
Although these effects were first considered more than half a century ago, our
understanding of these effects is still far from complete.
This is manifest in the fact that there is still very limited knowledge about
the influence of the external field's spacetime-dependence.
Besides numerical simulations (see, e.g.,~\cite{Gies2005, Dunne2006b, Ruf2009,
  Hebenstreit2009, *Hebenstreit2011a, Orthaber2011, Jiang2011, Jiang2012,
  Wollert2015a, Aleksandrov2016, *Aleksandrov2017}), several analytical methods
have been used for computing the pair production probability, such as the WKB
method~\cite{Popov1972, Marinov1977, Dumlu2011, Linder2015} or the worldline
instanton method~\cite{Dunne2005, *Dunne2006}.
However, most of the studies so far were limited to fields that depend on a
single coordinate, e.g.\ time~\cite{Popov1972, Marinov1977, DiPiazza2004,
  Dunne2005, *Dunne2006, Kleinert2008, Dumlu2011, Kim2011, Strobel2014,
  Linder2015}, a spatial coordinate~\cite{Kim2002, Dunne2005, *Dunne2006,
  Kleinert2008} or a light-cone coordinate~\cite{Tomaras2001, Fried2002,
  Avan2003, Hebenstreit2011b, Ilderton2014}; see also~\cite{Ilderton2015}.
Via the worldline instanton method, there have been a few works on truly
spacetime-dependent fields but these were limited to special
cases~\cite{Schneider2016, Dumlu2016, Torgrimsson2018} or a fully numerical
treatment, see, e.g.,~\cite{Schneider2018} (see also~\cite{Hebenstreit2010} for
a work using the Wigner formalism).
Regarding the WKB approach, there have been even less studies for background
fields depending on both space and time.

In this article, we present a WKB method based on the eikonal (or
Hamilton-Jacobi) equation that promises to overcome this fundamental restriction
(see~\cite{DiPiazza2014} for a previous approach to electron propagation based
on the eikonal equation).
For the sake of simplicity, we will consider the Dirac equation in 1+1
dimensions.
However, we believe that the method can be generalized to higher dimensions in a
straightforward way as long as the external field only depends on the time and a
single spatial coordinate.
As an important example, we shall study electron-positron pair creation due to a
spacetime-dependent mass $m(t,x)$ in the Dirac equation.
As one possible motivation, we note that a curved space-time metric such as in
cosmological particle production can be mapped to a spacetime-dependent mass in
the Dirac equation~\cite{Koke2016}.

The article is organized as follows:
We start by reviewing the conventional WKB method for the time-dependent Dirac
equation and use a specific time-dependent mass as an example in
Section~\ref{sec:wkb}.
In Section~\ref{sec:eikonalformalism}, we expand solutions of the Dirac equation
using solutions of the eikonal (or Hamilton-Jacobi) equation, giving two linear
coupled partial differential equations.
In Section~\ref{sec:knowncases} we show that these equations reduce to known
results if the electric field (or mass) either is purely time-dependent or
purely space-dependent.
Problems that occur while solving the eikonal equation with a truly
spacetime-dependent field are discussed in Section~\ref{sec:caustics}.
The case of a spacetime-dependent mass is considered in Section~\ref{sec:mass}.
We calculate approximative solutions to the equations mentioned above for a
spacetime-dependent mass with a weak space dependence in
Section~\ref{sec:weaklyspacedependent}.

\section{WKB formalism}%
\label{sec:wkb}

Let us start by briefly reviewing the standard derivation of the WKB formalism
for purely time-dependent fields in $1+1$ dimensions (see e.g.\ \cite{Dumlu2011,
  Linder2015} for comparison).
As we are interested in pair production due to a spacetime-dependent mass (or
scalar potential) later on, we consider the case of a time-dependent mass and a
time-dependent electric field in 1+1 dimensions.

We start with the covariant Dirac equation ($\hbar = c = 1$)
\begin{equation}
  \left[ i \gamma^\mu \left( \pd{\mu} + i q A_\mu \right) - m \right] \psi = 0,
  \label{eq:diraceq}
\end{equation}
where $A_\mu$ are the components of the electromagnetic potential and
$\gamma^\mu$ are the gamma matrices satisfying the Clifford algebra's
anticommutation relation
\begin{equation}
  \left\{ \gamma^\mu, \gamma^\nu \right\} = 2 \eta^{\mu \nu}.
\end{equation}
Now consider the Hamiltonian form of the Dirac equation in 1+1 dimensions in
temporal gauge $A_0 = 0$, $A_1 = A(t)$,
\begin{equation}
  i \pd{t} \psi(t, x) = \{ -i \gamma^0 \gamma^1 [\pd{x} + i q A(t)] + \gamma^0 m(t) \} \psi(t, x).
\end{equation}
After expanding $\psi(t, x)$ into Fourier modes $\psi_p(t)$ we get
\begin{equation}
  i \pd{t} \psi_p(t)
  = \{ \gamma^0 \gamma^1 [p + q A(t)] + \gamma^0 m(t) \} \psi_p(t) \!
  = \! H_p(t) \psi_p(t).
  \label{eq:wkbdiracfourier}
\end{equation}
Because $H_p^2(t) = m^2(t) + {[p + q A(t)]}^2 = \Omega_p^2(t)$, the self-adjoint
operator $H_p(t) = H_p^\dagger(t)$ has the instantaneous eigenvectors $u_\pm(p;
t)$,
\begin{equation}
  H_p(t) u_\pm(p; t) = \pm \Omega_p(t) u_\pm(p; t),
\end{equation}
which are orthonormal, i.e.\ $u_\pm^\dagger u_\pm = 1$ and $u_\pm^\dagger u_\mp
= 0$.
As usual, this normalization prescription still leaves the phases of the
spinors free to choose.
Additionally, one can show that
\begin{equation}
  \dot{u}_+^\dagger u_- = \co{\bigl( u_-^\dagger \dot{u}_+ \bigr)}
  = \frac{1}{2 \Omega_p} u_+^\dagger \dot{H}_p u_-, \qquad
  \dot{u}_\pm^\dagger u_\pm = 0.
\end{equation}
We expand $\psi_p(t)$ in terms of these eigenvectors
\begin{equation}
  \psi_p(t) = \alpha(p; t) u_+(p; t) e^{-i \varphi_p(t)} + \beta(p; t) u_-(p; t) e^{i \varphi_p(t)}
  \label{eq:psipexpansion}
\end{equation}
with the time-dependent phase (eikonal)
\begin{equation}
  \varphi_p(t) = \int_{-\infty}^t \!\!\!\!\!\! \diff{t'} \: \Omega_p(t').
  \label{eq:varphip}
\end{equation}
This expansion~\eqref{eq:psipexpansion} reflects the main idea of the WKB
approach, i.e., the separation of the rapid oscillation of the phases $\exp\{\pm
i\varphi_p(t)\}$ from the slow variation of the background in $u_\pm(p; t)$ as
well as $\alpha(p; t)$ and $\beta(p; t)$.

Upon inserting this expansion into~\eqref{eq:wkbdiracfourier} and projecting
onto $u_\pm(p; t)$, we get two coupled ordinary differential equations for
$\alpha(p; t)$ and $\beta(p; t)$,
\begin{equation}
  \begin{alignedat}{2}
    \dot{\alpha} &= &&\frac{\beta}{2 \Omega_p} e^{2i \varphi_p} u_+^\dagger \dot{H}_p u_-, \\
    \dot{\beta} &= -&&\frac{\alpha}{2 \Omega_p} e^{-2i \varphi_p} \co{\bigl( u_+^\dagger \dot{H}_p u_- \bigr)}.
  \end{alignedat}
\end{equation}
We define $\mathcal{R}(t) = \beta(p; t)/\alpha(p; t)$ and find a Riccati
equation
\begin{equation}
  \begin{aligned}
    \dot{\mathcal{R}} = -\frac{\Re (u_+^\dagger \dot{H}_p u_-)}{2 \Omega_p}
    &\bigl[ e^{-2i \varphi_p} + \mathcal{R}^2 e^{2i \varphi_p} \bigr] \\
    + \frac{i \Im (u_+^\dagger \dot{H}_p u_-)}{2 \Omega_p}
    &\bigl[ e^{-2i \varphi_p} - \mathcal{R}^2 e^{2i \varphi_p} \bigr].
  \end{aligned}
\end{equation}
Note that the exact form of the right-hand side depends on the chosen
representation and normalization of $u_+$ and $u_-$ due to the factor
$u_+^\dagger \dot{H}_p u_-$.
Using $\gamma^0 = \sigma_z$ and $\gamma^1 = i \sigma_y$ and assuming vanishing
phase difference between the spinors $u_{+}$ and $u_{-}$ we find
\begin{equation}
  \dot{\mathcal{R}} = \frac{m qE - (p + qA) \dot{m}}{2 \Omega_p^2}
  \bigl[ e^{-2i \varphi_p} + \mathcal{R}^2 e^{2i \varphi_p} \bigr]
  \label{eq:riccatieq}
\end{equation}
which for $\dot{m} = 0$ reduces to the well-known form
\begin{equation}
  \dot{\mathcal{R}} = \frac{m qE}{2 \Omega_p^2} \bigl[ e^{-2i \varphi_p} + \mathcal{R}^2 e^{2i \varphi_p} \bigr].
\end{equation}
On the other hand, for $A(t) = 0$ we find
\begin{equation}
  \dot{\mathcal{R}} = -\frac{p \dot{m}}{2 \Omega_p^2} \bigl[ e^{-2i \varphi_p} + \mathcal{R}^2 e^{2i \varphi_p} \bigr].
  \label{eq:riccatieqm}
 \end{equation}
The number of created positrons (or electrons) with momentum $p$ can be
calculated using (see Appendix~\ref{sec:pairproduction})
\begin{equation}
  N_{e^+}(p) \propto \abs{\beta_{\mathrm{out}}(p)}^2
  = \frac{\abs{\mathcal{R}_{\mathrm{out}}}^2}{1 + \abs{\mathcal{R}_{\mathrm{out}}}^2}
\end{equation}
where $\beta_{\mathrm{out}}(p) = \beta(p; t \to \infty)$,
$\mathcal{R}_{\mathrm{out}} = \mathcal{R}(t \to \infty)$ and we have used
the relation $\abs{\alpha}^2 + \abs{\beta}^2 = 1$ in the last equality.
Under the assumption that few pairs are created, i.e.\ $\mathcal{R} \ll 1$, a
linearized form of the Riccati equation~\eqref{eq:riccatieq} is often used:
\begin{equation}
  \dot{\mathcal{R}}(t) \approx
  \frac{m qE - (p + qA) \dot{m}}{2 \Omega_p^2} e^{-2i \varphi_p}.
  \label{eq:riccatieqlinear}
\end{equation}
In that case we get $N_{e^+}(p) \propto \abs{\mathcal{R}_{\mathrm{out}}}^2$.
To obtain $\mathcal{R}_{\mathrm{out}}$ we integrate the linearized Riccati
equation~\eqref{eq:riccatieqlinear} over all times
\begin{equation}
  \mathcal{R}_{\mathrm{out}} \approx \int_{-\infty}^{\infty} \diff{t}
  \frac{m qE - (p + qA) \dot{m}}{2 \Omega_p^2} e^{-2i \varphi_p}.
  \label{eq:Rpoutintegral}
\end{equation}
For symmetric electric fields $A(-t) = -A(t)$ with a constant mass ($\dot{m} =
0$) one expects the maximum number of created pairs for $p = 0$ as the
denominator of the integrand is minimal in that case.
On the other hand, in the case with only a time-dependent mass ($A(t) = 0$) the
right-hand side of~\eqref{eq:riccatieqm} immediately reveals that
$\mathcal{R}(t)$ vanishes for $p = 0$ and so does the number of produced pairs.

Furthermore, upon deforming the integration contour for the
integral~\eqref{eq:Rpoutintegral} in the complex plane we see that the integrand
is exponentially suppressed in the lower half-plane.
Thus, the integral's value is dominated by the value of the exponential at the
singularity closest to the real axis.
This singularity at $t_{*}$ could be a pole of the prefactor $\Omega_{p}(t_{*})
= 0$ or a branch point or any other point where the integrand is not analytic
anymore and thus we cannot deform the integration contour further.
Then, $\mathcal{R}_{\mathrm{out}}$ can be approximated as
\begin{equation}
  \mathcal{R}_{\mathrm{out}} \sim e^{-2i \varphi_{p}(t_{*})}.
\end{equation}
This estimate does not give the correct prefactor but only the exponent.
However, due to the linearization of the Riccati equation one cannot
realistically expect to obtain the prefactor from the
integral~\eqref{eq:Rpoutintegral} exactly anyway.
If there are multiple singularities that are comparably close to the real axis,
contributions from all singularities have to be taken into account which leads
to interference effects in the momentum spectrum~\cite{Dumlu2011, Orthaber2011,
  Fey2012}.

\subsection{Example: time-dependent mass}

As an example, we want to calculate the number of produced pairs for a specific
time-dependent mass as a toy model.
We will use a similar functional dependence later in
Section~\ref{sec:weaklyspacedependent} as a spacetime-dependent mass where some
of the results derived here will be useful.

We use a mass of the form
\begin{equation}
  m(t) = m_{0} \sqrt{1 + {\biggl[ \frac{f(\omega t)}{\gamma} \biggr]}^{2}}
  \label{eq:timedependentmass}
\end{equation}
with $f(\tau) = \sech \tau$ and a dimensionless parameter $\gamma$ that controls
the amplitude of the pulse.
For large $\gamma$, the relative change of $m(t)$ is small and we may use
perturbation theory to estimate the pair-creation probability (see below).
For small $\gamma$, however, the change is large and we need another method,
such as the WKB approach.

This parameter $\gamma$ also controls the adiabaticity, i.e., the applicability
of the WKB approximation.
A measure for the adiabaticity is the rate of change $\dot m$ of the mass
compared to the mass itself, i.e., $\dot m/m^2$, which scales with
$\gamma\omega/m_0$.
Thus, the interesting region of small $\gamma$ can be treated via the WKB
approach provided that $\omega\ll m_0$.

The expression~\eqref{eq:timedependentmass} is motivated by the fact that
typically the squares of mass (or potential) terms are added.
As an example, let us consider the Dirac equation in 2+1 dimensions where the
second spatial dimension is compactified, giving rise to a discrete Kaluza-Klein
tower of transversal momenta $k_\perp$.
Then, the effective masses of the 1+1 dimensional Dirac equations would be
$m_{\rm 1D}^2=k_\perp^2+m^2_{\rm 2D}$.
As another example, let us consider a scalar field $(\Box+m^2)\phi+V'(\phi)=0$
with the interaction potential $V(\phi)$.
Then, linearization $\phi=\phi_0+\delta\phi$ around a given background solution
$\phi_0$ yields the effective mass $m^2_{\rm eff}=V''(\phi_0)+m^2$ for the
perturbation $\delta\phi$.

The parameter $\gamma$ plays a role very analogous to the Keldysh parameter
$\gamma=m\omega/(qE)$ for strong electric fields, see, e.g., \cite{Popov1971}.
This analogy can be made even more explicit by considering the form of the
effective mass $m_\mathrm{eff} = m \sqrt{1 - \braket{qA_{\mu} qA^{\mu}}/m^{2}}$
of an electron within a laser pulse (see~\cite{Kibble1966, Dodin2008,
  Kohlfurst2014}), even though the resulting pair-creation probability should be
derived by using the vector potentials $A_\mu$ directly.

We then can approximate $\mathcal{R}_{\mathrm{out}}$ using the linearized
Riccati equation~\eqref{eq:riccatieqlinear}
\begin{equation}
  \mathcal{R}_{\mathrm{out}} \approx -\frac{1}{\tilde{\gamma}^{2}} \frac{p}{m_{0}}
  \int_{-\infty}^{\infty} \diff{\tau} \frac{f(\tau) f'(\tau) e^{-2i \varphi_{p}(\tau)}}
      {\sqrt{1 + {\Bigl[ \frac{f(\tau)}{\gamma} \Bigr]}^{2}}
        \biggl\{ 1 + {\Bigl[ \frac{f(\tau)}{\tilde{\gamma}} \Bigr]}^{2} \biggr\}}
  \label{eq:Rpoutmass}
\end{equation}
where
\begin{equation}
  \varphi_{p} = \frac{m_{0}}{\omega} \sqrt{1 + {\Bigl( \frac{p}{m_{0}} \Bigl)}^{2}}
  \int_{-\infty}^{\tau} \!\! \diff{\tau'} \sqrt{1 + {\biggl[ \frac{f(\tau')}{\tilde{\gamma}} \biggr]}^{2}}.
  \label{eq:varphipmass}
\end{equation}
and
\begin{equation}
  \tilde{\gamma} = \gamma \sqrt{1 + {\Bigl( \frac{p}{m_{0}} \Bigl)}^{2}}.
\end{equation}
For $f(\tau) = \sech \tau$ the phase integral can be calculated analytically,
giving
\begin{equation}
  \varphi_{p} = \frac{m_{0}}{\omega} \sqrt{1 + {\Bigl( \frac{p}{m_{0}} \Bigl)}^{2}}
  \bigl[ \phi(\tau) - \phi(-\infty) \bigr]
\end{equation}
where
\begin{equation}
  \begin{aligned}
    \phi(\tau) = \frac{1}{\tilde{\gamma}}
    &\arctan \Biggl[ \frac{\sinh \tau}{\sqrt{1 + \tilde{\gamma}^{2} \cosh^{2} \tau}} \Biggr] \\
    + &\artanh \Biggl[ \frac{\tilde{\gamma} \sinh \tau}{\sqrt{1 + \tilde{\gamma}^{2} \cosh^{2} \tau}} \Biggr].
  \end{aligned}
\end{equation}
The integral for $\mathcal{R}_{\mathrm{out}}$ is dominated by the value of
the exponent at the pole where $f(\tau_{*}) = \pm i \tilde{\gamma}$,
\begin{equation}
  \abs{\mathcal{R}_{\mathrm{out}}}^{2}
  \sim \abs{e^{-2i \varphi_{p}(\tau_{*})}}^{2}
  = e^{4 \Im \varphi_{p}(\tau_{*})}.
\end{equation}
For $f(\tau) = \sech \tau$ we find
\begin{equation}
  \tau_{*} = \arcosh\biggl( \pm \frac{i}{\tilde{\gamma}} \biggr)
  = \ln \Biggl[ \frac{1}{\abs{\tilde{\gamma}}}
    + \sqrt{1 + \biggl( \frac{1}{\tilde{\gamma}^{2}} \biggr)} \Biggr]
  - i \frac{\pi}{2}
  \label{eq:taustar}
\end{equation}
and thus
\begin{equation}
  \abs{\mathcal{R}_{\mathrm{out}}}^{2} \sim \exp \Biggl[
  -2 \pi \frac{m_{0}}{\omega} \sqrt{1 + {\Bigl( \frac{p}{m_{0}} \Bigl)}^{2}}
  \Biggr].
  \label{eq:sechmassRpanalytic}
\end{equation}
This result does not depend on $\gamma$ which at first is a bit surprising.
E.g.\ in the limit $\gamma \to \infty$, $m(t) = m_{0} = \text{const.}$ and thus
no pairs should be produced.
This apparent inconsistency can be resolved by the observation that our WKB
approach breaks down for large enough $\gamma$, where we should use perturbation
theory instead (see above).

To confirm our result we computed $\mathcal{R}_{\mathrm{out}}$ numerically
from the full Riccati equation~\eqref{eq:riccatieqm}.
Due to the highly-oscillatory coefficients in the Riccati equation we integrated
the equation using the \texttt{TIDES} library~\cite{Abad2012} in conjunction
with the arbitrary-precision library \texttt{MPFR}~\cite{Fousse2007}.
To parallelize the computation, \texttt{GNU Parallel}~\cite{Tange2011} has been
used.

\begin{figure}
  \centering
  \includegraphics{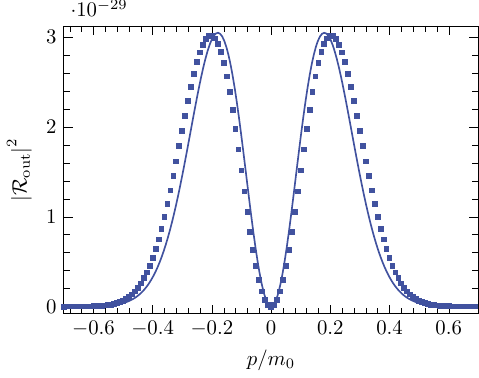}
  \caption{\label{fig:sechmassRp}Plot of the density of produced pairs
    $\abs{\mathcal{R}_{\mathrm{out}}}^{2}$ for the $\sech$-mass
    in~\eqref{eq:timedependentmass} where $f(\tau) = \sech \tau$ with $\omega =
    0.1 m_{0}$ and $\gamma = 0.1$. The squares are numerically calculated results
    while the solid line represents the analytical estimate
    from~\eqref{eq:sechmassRpanalytic}. We used $a p^{2}$ as the prefactor of
    the analytical result with $a = 5/m_{0}^{2}$ chosen to fit the height of the
    peaks in the numerical result; see~\eqref{eq:Rpoutintegral}.}
\end{figure}
Figure~\ref{fig:sechmassRp} shows the analytical result
from~\eqref{eq:sechmassRpanalytic} and the numerical result for
$\abs{\mathcal{R}_{\mathrm{out}}}^{2}$ together for a specific choice of
$\gamma$ and $\omega$.
Because the approximation in~\eqref{eq:sechmassRpanalytic} does not produce the
correct prefactor we assume it to be $a p^{2}$.
This is motivated by the form of the integrand's prefactor
in~\eqref{eq:Rpoutintegral} which for $E = 0$ is proportional to the canonical
momentum $p$.
The constant $a$ is then chosen to fit the numerical data.

We find very good agreement between the analytical estimate and the numerical
calculation.
Even without the heuristically determined factor $a$ the analytic approximation
lies within an order of magnitude of the numerical result.

\begin{figure}
  \includegraphics{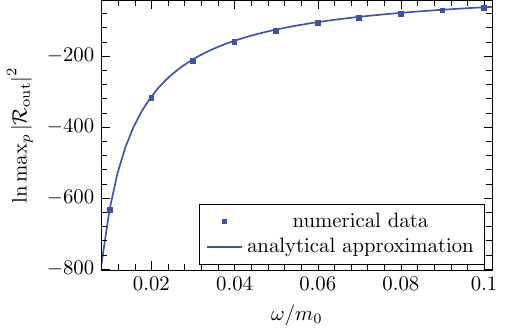}
  \caption{\label{fig:sechmassOmega}Plot of the logarithm of the maximum of
    $\abs{\mathcal{R}_{\mathrm{out}}}^{2}$ for different values of $\omega$
    and $\gamma = 0.1$. The plot shows both the numerical results and the
    analytical approximation from~\eqref{eq:sechmassRpanalytic}.}
\end{figure}

\begin{figure}
  \includegraphics{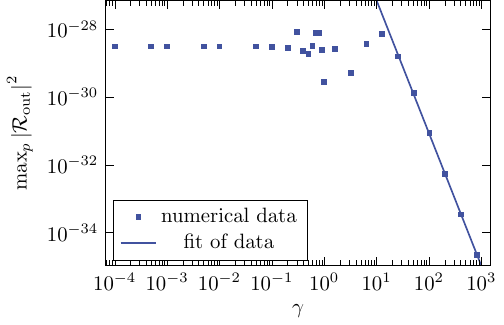}
  \caption{\label{fig:sechmassGamma}Log-log plot of the maximum of
    $\abs{\mathcal{R}_{\mathrm{out}}}^{2}$ for different values of $\gamma$ and
    $\omega = 0.1 m_{0}$. The plot shows both the numerical results and a fit of
    the numerical data for large $\gamma$. The slope of the fitted line is
    $-3.91$. We see the crossover from the non-perturbative WKB
    regime~\eqref{eq:sechmassRpanalytic} for small $\gamma$ to the perturbative
    regime $\sim\gamma^{-4}$ for large $\gamma$.}
\end{figure}

Indeed, if one plots the values of the numerical results' peaks over different
values of $\omega$, the points fall nicely on the curve predicted by the maximum
of the exponential in~\eqref{eq:sechmassRpanalytic} (see
Fig.~\ref{fig:sechmassOmega}).
On the other hand, if we fix $\omega$ and vary $\gamma$ the maximum of the
numerical data behaves as in Fig.~\ref{fig:sechmassGamma}.
For small $\gamma \ll 1$ the maximum remains constant while for large $\gamma
\gg 1$ the maximum seems to go like $\gamma^{-4}$.
This behavior is due to the prefactor in~\eqref{eq:Rpoutmass} which goes like
$\gamma^{-2}$ for large $\gamma \gg 1$.
In between these two regions the value of the maximum fluctuates.
This can be attributed to the prefactor as well because the order of magnitude
does not change as one would expect if this behavior came from the exponent.
%

\section{Eikonal formalism}%
\label{sec:eikonalformalism}

We now want to develop a more general procedure for calculating the pair
production probability that in principle also works for spacetime-dependent
fields.
The main idea of the WKB formalism as presented in the last section is to
separate fast and slow oscillations in the wave function:
The factor of $\exp[ \pm i \varphi_{p}(t) ]$ contains the fast oscillations
while the prefactors $\alpha$ and $\beta$ contain the slow oscillations.
We try a similar approach for spacetime-dependent fields.

First, we define two operators
\begin{equation}
  M_\pm = - \gamma^\mu \left( \pd{\mu} S_\pm + q A_\mu \right)
  = - \gamma^\mu \Pi_\mu^\pm,
\end{equation}
with $S_\pm$ being the two independent solutions of the relativistic eikonal (or
Hamilton-Jacobi) equation
\begin{equation}
  \eta^{\mu \nu} \left( \pd{\mu} S_\pm + q A_\mu \right)
            \left( \pd{\nu} S_\pm + q A_\nu \right)
  = \eta^{\mu \nu} \Pi_\mu^\pm \Pi_\nu^\pm
  = m^2.
  \label{eq:eikonaleq}
\end{equation}
The eikonal equation above can be obtained from classical electrodynamics.
Thus, it could be derived from the Dirac equation~\eqref{eq:diraceq} via
inserting the WKB ansatz $\psi\sim\exp(i S_{\pm}/\hbar)$ and only keeping the
lowest-order terms in $\hbar$.
However, here we motivate the WKB expansion by assuming that the mass $m$ is the
largest relevant scale in our problem, leading to rapid oscillations of $\exp(i
S_{\pm}/\hbar)$.

We will use the convention that $S_+$ and $S_-$ correspond to solutions with
positive and negative energy, respectively,
\begin{equation}
  \Pi_t^\pm = \pd{t} S_\pm + q A_0
  = \mp \sqrt{m^2 + {\bigl( \nabla S_\pm + q \vect{A} \bigr)}^2}.
\end{equation}
Note that this eikonal equation is an immediate generalization
of~\eqref{eq:varphip}.
When $A_\mu$ and $m$ are constant, the solutions $S_\pm$ correspond to
plane wave solutions, that is $S_\pm = \mp p_\mu x^\mu$.

Squaring the operators $M_\pm$, we get
\begin{equation}
  M_\pm^2
  = \gamma^\mu \gamma^\nu \Pi_\mu^\pm \Pi_\nu^\pm
  = \eta^{\mu \nu} \Pi_\mu^\pm \Pi_\nu^\pm
  = m^2.
\end{equation}
Thus, the operators $M_\pm$ both have the two distinct eigenvalues $\pm m$.
Let $u_\pm$ and $v_\pm$ be their respective eigenvectors defined as follows
\begin{equation}
  M_+ u_\pm = \pm m u_\pm, \quad M_- v_\pm = \pm m v_\pm.
\end{equation}
Because the operators $M_\pm$ are self-adjoint in the sense that $\dadj{M}_\pm =
\gamma^{0} M_{\pm}^{\dagger} \gamma^{0} = M_\pm$, their eigenvectors are
orthogonal,
\begin{equation}
  \dadj{u}_{+} u_{-} = \dadj{u}_{-} u_{+} = \dadj{v}_{+} v_{-} = \dadj{v}_{-} v_{+} = 0,
\end{equation}
where $\dadj{u}_{\pm} = u_{\pm}^{\dagger} \gamma^{0}$ and analogously for
$\dadj{v}_{\pm}$.
We normalize the eigenvectors as follows:
\begin{equation}
  \dadj{u}_{+} u_{+} = -\dadj{u}_{-} u_{-} = -\dadj{v}_{+} v_{+} = \dadj{v}_{-} v_{-} = 1.
\end{equation}
Although parts of the following derivation can be carried out in a general
manner, we want to focus on the case of a 1+1-dimensional spacetime.
Then it is sufficient to use $2 \times 2$ matrices for the gamma matrices and
the $M_\pm$ will only have one eigenvector each for every eigenvalue.
We expand the spinor $\psi$ in terms of these eigenvectors,
\begin{equation}
  \psi = \alpha u_+ e^{i S_+} + \beta v_+ e^{i S_-},
  \label{eq:eikonalexpansion}
\end{equation}
which is motivated by the expansion~\eqref{eq:psipexpansion} of the spinor in
the time-dependent case.
There, the functions $\alpha$ and $\beta$ were the Bogoliubov coefficients of
the transformation between in- and out-states (see
Appendix~\ref{sec:pairproduction}) and therefore we will sometimes refer to them
as Bogoliubov coefficients here as well.
Using the expansion~\eqref{eq:eikonalexpansion}, the Dirac
equation~\eqref{eq:diraceq} reduces to
\begin{equation}
  i \gamma^\mu \pd{\mu} \left( \alpha u_+ \right) e^{i S_+}
  + i \gamma^\mu \pd{\mu} \left( \beta v_+ \right) e^{i S_-} = 0.
  \label{eq:diraceqaftereikonalexpansion}
\end{equation}
In terms of the large-$m$ or small-$\hbar$ expansion mentioned after the eikonal
equation~\eqref{eq:eikonaleq}, the leading-order contribution gives
Eq.~\eqref{eq:eikonaleq} for the exponent $S_\pm$, while the sub-leading order
determines the above equation for the Bogoliubov coefficients, compare Eq.~(2)
in~\cite{Note1}.

Multiplying~\eqref{eq:diraceqaftereikonalexpansion} by $\dadj{u}_+$ or
$\dadj{v}_+$ from the left we get two coupled partial differential equations
\begin{equation}
  \begin{alignedat}{3}
    &\dadj{u}_+ \gamma^\mu \pd{\mu} \alpha u_+
    &&= -\dadj{u}_+ \gamma^\mu (\pd{\mu} \beta v_+) \: &&e^{-i (S_+ - S_-)}, \\
    &\dadj{v}_+ \gamma^\mu \pd{\mu} \beta v_+
    &&= -\dadj{v}_+ \gamma^\mu (\pd{\mu} \alpha u_+) \: &&e^{i (S_+ - S_-)}.
  \end{alignedat}
  \label{eq:diraceqproj}
\end{equation}
Analogous to the Dirac convention, we choose
\begin{equation}
  \gamma^0 = \sigma_z = \begin{pmatrix} 1 & 0 \\ 0 & -1 \end{pmatrix}, \quad
  \gamma^1 = i \sigma_x = \begin{pmatrix} 0 & i \\ i & 0 \end{pmatrix}
\end{equation}
for the gamma matrices. Thus,
\begin{equation}
  M_\pm =
  \begin{pmatrix}
    -\Pi_t^\pm & -i\Pi_x^\pm \\
    -i\Pi_x^\pm & +\Pi_t^\pm
  \end{pmatrix}
\end{equation}
and the eigenvectors $u_\pm$ and $v_\pm$ can be written as
\begin{equation}
  \begin{alignedat}{2}
    u_+ &= N_+ \begin{pmatrix} m - \Pi_t^+ \\ -i \Pi_x^+ \end{pmatrix},
    u_- &&= N_+ \begin{pmatrix} i \Pi_x^+ \\ m - \Pi_t^+ \end{pmatrix}
    = i \gamma^0 \gamma^1 u_+, \\
    v_+ &= N_- \begin{pmatrix} -i \Pi_x^- \\ m + \Pi_t^- \end{pmatrix},
    v_- &&= N_- \begin{pmatrix} m + \Pi_t^- \\ i \Pi_x^- \end{pmatrix}
    = i \gamma^0 \gamma^1 v_+,
  \end{alignedat}
  \label{eq:uveigenvectors}
\end{equation}
with the normalization constants
\begin{equation}
  N_\pm = \frac{1}{\sqrt{2m (m \mp \Pi_t^\pm)}}.
\end{equation}
After calculating all the inner products that appear in~\eqref{eq:diraceqproj}
we get the following equations for $\alpha$ and $\beta$:
\begin{equation}
  \begin{aligned}
    &\eta^{\mu \nu} \Pi_\mu^+ \pd{\nu} \alpha
    - \frac{1}{2m^2} \bigl(
    \eta^{\mu \rho} \eta^{\nu \lambda} - \eta^{\mu \nu} \eta^{\rho \lambda}
    \bigr) \Pi_\rho^+ \Pi_\lambda^+ (\pd{\mu} \Pi_\nu^+) \alpha \\
    &= i m e^{-i(S_+ - S_-)} \biggl[
      \kappa^\mu \pd{\mu} \beta
      + \frac{1}{2 m^2}
      \varepsilon^{\lambda \nu} \varepsilon\indices{_\rho^\mu} \kappa^\rho
      \Pi_\lambda^- (\pd{\mu} \Pi_\nu^-) \beta
      \biggr], \\[1ex]
    &\eta^{\mu \nu} \Pi_\mu^- \pd{\nu} \beta
    - \frac{1}{2m^2} \bigl(
    \eta^{\mu \rho} \eta^{\nu \lambda} - \eta^{\mu \nu} \eta^{\rho \lambda}
    \bigr) \Pi_\rho^- \Pi_\lambda^- (\pd{\mu} \Pi_\nu^-) \beta \\
    &= i m e^{i(S_+ - S_-)} \biggl[
      \kappa^\mu \pd{\mu} \alpha
      - \frac{1}{2 m^2}
      \varepsilon^{\lambda \nu} \varepsilon\indices{_\rho^\mu} \kappa^\rho
      \Pi_\lambda^+ (\pd{\mu} \Pi_\nu^+) \alpha
      \biggr],
  \end{aligned}
  \label{eq:alphabetaexact}
\end{equation}
where
\begin{equation}
  \kappa^\mu 
  = N_+ N_-
  \begin{pmatrix}
    \Pi_x^+ (m + \Pi_t^-) - \Pi_x^- (m - \Pi_t^+)  \\
    (m - \Pi_t^+) (m + \Pi_t^-) - \Pi_x^+ \Pi_x^-
  \end{pmatrix}.
  \label{eq:kappa}
\end{equation}
These equations~\eqref{eq:alphabetaexact} are completely equivalent to the Dirac
equation~\eqref{eq:diraceq}, but might offer advantages for numerical
simulations and for analytical approximations (see below).
For the purely time-dependent case, it is known that solving the quantum kinetic
equations (see, e.g.,~\cite{Hebenstreit2009, Hebenstreit2010}) or the Riccati
equation (see, e.g.,~\cite{Popov1972, Marinov1977, Dumlu2011, Linder2015}) can
be more efficient numerically than the original Dirac equation.
Thus, we expect that similar advantages could apply here, especially in cases
where the functions $S_\pm$ are available analytically (e.g., within suitable
approximations) or can be efficiently implemented numerically.

\section{Known limiting cases}%
\label{sec:knowncases}

We now want to show that the equations~\eqref{eq:alphabetaexact} reproduce the
correct results for both a time-dependent electric field with a time-dependent
mass and a space-dependent electric field.

\subsection{Time-dependent electric field and mass}

We use the temporal gauge where
\begin{equation}
  A_0 = 0, \quad
  A_1 = A(t), \quad
  E = \dot{A}(t).
\end{equation}
Then the two independent solutions of the eikonal equation~\eqref{eq:eikonaleq}
are given by
\begin{equation}
  S_\pm = \mp \varphi_p(t) + px
  \label{eq:Spmtimedependent}
\end{equation}
with $\varphi_p(t)$ as in Sec.~\ref{sec:wkb}.
We thus find
\begin{gather}
  \Pi_t^\pm = \mp \Omega_p (t), \quad
  \Pi_x^\pm = p + q A(t), \\
  N_\pm = \frac{1}{\sqrt{2m(t) [m(t) + \Omega_p(t)]}}, \quad
  \kappa^\mu = \begin{pmatrix} 0 \\ 1 \end{pmatrix}.
\end{gather}
None of the coefficients in~\eqref{eq:alphabetaexact} depends on $x$ in this
case.
Thus, if we impose boundary conditions such that $\alpha$ and $\beta$ are
constant initially (i.e.\ for $t \to -\infty$) then $\pd{x} \alpha = \pd{x}
\beta = 0$ for all times.
The equations~\eqref{eq:alphabetaexact} for $\alpha$ and $\beta$ then simplify
to
\begin{equation}
  \begin{aligned}
    \Omega_p \pd{t} \alpha
    + \frac{1}{2} \dot{\Omega}_p \alpha
    = -\frac{i}{2} \frac{m qE - (p + qA) \dot{m}}{\Omega_p} \beta e^{2i \varphi_p}, \\
    \Omega_p \pd{t} \beta
    + \frac{1}{2} \dot{\Omega}_p \beta
    = \frac{i}{2} \frac{m qE - (p + qA) \dot{m}}{\Omega_p} \alpha e^{-2i \varphi_p}.
  \end{aligned}
  \label{eq:alphabetatimedependent}
\end{equation}
We define the ratio $\mathcal{R}(t) = \beta(t)/\alpha(t)$ and,
using~\eqref{eq:alphabetatimedependent}, calculate its time derivative
\begin{equation}
  \begin{aligned}
    \pd{t} \mathcal{R} &= \frac{\pd{t} \beta}{\alpha}
    - \mathcal{R}^2 \frac{\pd{t} \alpha}{\beta} \\
    &= i \frac{m qE - (p + qA) \dot{m}}{2 \Omega_p^2} \bigl[
      e^{-2i \varphi_p} + \mathcal{R}^2 e^{2i \varphi_p}
      \bigr],
  \end{aligned}
\end{equation}
which is a Riccati equation that is up to an factor of $i$ (that can be
attributed to a different normalization for the spinors $u_{+}$ and $v_{+}$ used
here than for the spinors $u_{\pm}$ in Sec.~\ref{sec:wkb}) identical to the one
in ordinary time-dependent WKB (compare~\eqref{eq:riccatieq}).

\subsection{Space-dependent electric field}

For a purely space-dependent electric field (compare~\cite{Nikishov1970,
  Kim2002, Kleinert2008}) we use the gauge
\begin{equation}
  A_0 = \phi(x), \quad A_1 = 0, \quad E = -\phi'(x).
\end{equation}
In complete analogy to the time-dependent case, we find
\begin{equation}
  S_\pm = -\omega t \pm \varphi_\omega(x),
\end{equation}
with
\begin{equation}
  \varphi_\omega(x) = \int_{-\infty}^x \!\!\!\!\!\! \diff{x'} \: P_\omega(x'), \quad
  P_\omega(x) = \sqrt{{\bigl[ \omega - q \phi(x) \bigr]}^2 - m^2}.
\end{equation}
Thus
\begin{gather}
  \Pi_t^\pm = -\omega + q \phi(x), \quad
  \Pi_x^\pm = \pm P_\omega(x), \\
  N_\pm = \frac{1}{\sqrt{2m [m \pm (\omega - q \phi(x))]}}, \quad
  \kappa^\mu = \begin{pmatrix} -i \\ 0 \end{pmatrix}.
\end{gather}
Again, the coefficients in the equations for $\alpha$ and
$\beta$~\eqref{eq:alphabetaexact} are solely space-dependent and by requiring
that $\alpha$ and $\beta$ are constant left of the barrier, i.e.\ for $x \to
-\infty$ we find that $\pd{t} \alpha = \pd{t} \beta = 0$ for all values of $x$.
Then, after introducing the ratio $\mathcal{R} = \beta/\alpha$ we again find a
Riccati equation,
\begin{equation}
  \pd{x} \mathcal{R} = -\frac{m qE(x)}{2 P_\omega^2(x)} \bigl[
    e^{2i \varphi_\omega(x)} + \mathcal{R}^2 e^{-2i \varphi_\omega(x)}
    \bigr].
\end{equation}
This case is related to the one-dimensional Schr\"odinger scattering problem
from non-relativistic quantum mechanics.
Again the WKB expansion~\eqref{eq:eikonalexpansion} is motivated by the
separation of the rapidly oscillating phase $\exp\{iS_\pm\}$ from the slowly
varying rest.
This assumes that the local momentum scale $P_\omega(x)$ is much larger than all
other relevant scales, such as $P_\omega^2(x)\gg|P_\omega'(x)|$ or equivalently
$(\pd{x} S_{\pm})^{2}\gg\abs{\pd[2]{x} S_{\pm}}$.
Of course, this assumption breaks down at the classical turning points where
$P_{\omega}(x) = 0$.
Even though the above Riccati equation is in principle exact, integrating it
becomes problematic at those points.
Note that, in contrast to the purely time-dependent case, these classical
turning points $x$ can be real for sub-barrier tunneling problems.
For quantum reflection above the barrier, they are again complex.

\section{Caustics}%
\label{sec:caustics}

If we consider a truly spacetime-dependent problem difficulties in solving the
eikonal equation~\eqref{eq:eikonaleq} may occur.
Due to the non-linear nature of the eikonal equation, it may not be possible to
find global solutions in a classical sense, i.e.\ a solution might not be
differentiable everywhere.
Note that these singularities of the eikonal equation~\eqref{eq:eikonaleq} do
not (necessarily) imply that the solutions of the original Dirac
equation~\eqref{eq:diraceq} become singular.
They just indicate that the lowest-order WKB
approach~\eqref{eq:eikonalexpansion} employed here breaks down.
This is very similar to caustics in geometric (ray) optics -- e.g., the rainbow
effect -- where the density of light rays shows a singularity while the full
solution of the wave equation remains perfectly regular.
Another example is the one-dimensional stationary Schr\"odinger scattering
problem (discussed above) where the WKB approach breaks down at the classical
turning points (indicating the onset of tunneling) while the solutions to the
original Schr\"odinger equation remain perfectly regular.

We use the method of characteristics to visualize such situations (see
e.g.\ \cite{Evans1998} or many other standard text books on partial differential
equations for more details).
Using this method any first-order partial differential equation can be cast as a
system of ordinary differential equations by finding certain characteristic
curves along which the solution of the partial differential equation can be
integrated easily.
Afterwards, the solutions along multiple of those curves can be combined into a
solution surface.
This essentially amounts to going over to another set of coordinates where one
coordinate is the parameter to move along the curve and the other coordinates
number the curves.

Difficulties appear where two characteristic curves intersect.
At such a point the solution is not uniquely defined as we might use the value
on either one of the intersecting characteristics.
Multiple of those points form a caustic surface.

\begin{figure}
  \includegraphics{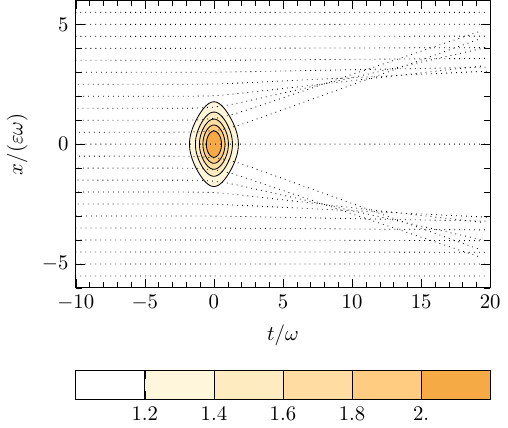}
  \caption{\label{fig:sechmasscharacteristics}Projected characteristic curves
    (dotted) for the $m(t, x)$ from~\eqref{eq:weaklyspacedependentmass}
    together with $m(t, x)$ itself (contour) using $f(\tau) = \sech \tau$,
    $g(\chi) = \sech \chi$, $m_0 = 1$, $p = 0$, $\omega = 0.8$, $\varepsilon =
    0.375$ and $\gamma = 0.5$.}%
\end{figure}

For example, Fig.~\ref{fig:sechmasscharacteristics} shows the
spacetime-dependent mass given in~\eqref{eq:weaklyspacedependentmass} together
with the (numerically calculated) characteristic curves.
We see that such a spacetime-dependent mass has a focusing/defocusing effect on
the characteristic curves similar to optical lenses on light rays.
Indeed we can estimate that the onset of the caustic surface is at time
\begin{equation}
  t_{f} \sim \frac{\gamma^{2} \omega}{\varepsilon^{2} \omega^{2}},
  \label{eq:caustictf}
\end{equation}
for $p = 0$ and $m$ only weakly space-dependent; see
Appendix~\ref{sec:causticsestimation} for details.

Pictures like Fig.~\ref{fig:sechmasscharacteristics} are well known from
geometrical optics.
In fact, geometrical optics is just an approximation to wave optics based on the
eikonal equation (for optics).
That is why it is not too surprising that the above formula~\eqref{eq:caustictf}
for $p = 0$ is strikingly similar to the formula for the focal length of a thin,
biconvex spherical lens~\cite{Hecht1979}
\begin{equation}
  f \propto \frac{L^2}{D (n_2 - n_1)},
\end{equation}
where $L \sim 1/(\varepsilon \omega)$, $D \sim 1/\omega$ and $n_2 - n_1 \sim
1/\gamma^{2}$.
We see that when the spatial inhomogeneity is weak (i.e.\ $\varepsilon$ small)
the caustics occur far away from the spacetime region in which the mass is
non-constant, i.e.\ where pairs are produced.
Thus, in the case of a purely time-dependent problem no caustics occur and our
solution is differentiable everywhere (compare~\eqref{eq:Spmtimedependent}).

In conclusion, assuming that the space-time region of particle creation is
sufficiently localized and that the spatial dependence is weak enough (compared
to the temporal variation), the potential problem of caustics (indicating
singular solutions of the eikonal equation) occurs far away from the space-time
region where the particles are created and thus does not invalidate our
analysis.
To cast this statement in a more formal form, one could take two routes:
As one option, one could choose a finite final time $t_{\rm out}$ which is large
enough such that it occurs after all pair-creation processes have taken place,
but small enough such that it is still before any caustics appear.
As another option, one could apply a mild deformation of the mass function
$m(t,x)$ in this time window which is so slow that the generated pair creation
(i.e., mixing of positive and negative frequencies) can be neglected, but which
undoes the focusing/defocusing effects and thus avoids caustics.

\section{Spacetime-dependent mass}%
\label{sec:mass}

We now want to turn to a truly spacetime-dependent problem, namely that of a
spacetime dependent mass $m(t, x)$ with no electromagnetic potential,
i.e.\ $A_\mu = 0$.
That case occurs in a 1+1-dimensional spacetime with curvature:
Every 1+1-dimensional spacetime is conformally flat, i.e.\ its metric
can be written as
\begin{equation}
  \diff{s}^2 = \mho^2(t, x) ( \diff{t}^2 - \diff{x}^2 ).
\end{equation}
Writing down the Dirac equation in such a spacetime reveals that it is
equivalent to the Dirac equation in flat space-time, but with a
spacetime-dependent mass $m(t, x) = \mho(t, x) m_0$ (see e.g.~\cite{Koke2016}
for details).

In that case, the eikonal equation~\eqref{eq:eikonaleq} is considerably simpler:
\begin{equation}
  \eta^{\mu \nu} \bigl( \pd{\mu} S_\pm \bigr) \bigl( \pd{\nu} S_\pm \bigr) = m^2(t, x).
  \label{eq:eikonaleqmass}
\end{equation}
We may write the two independent solutions $S_+$ and $S_-$ using two different
functions $R$ and $S$ by splitting $S_\pm$ into a symmetric and an antisymmetric
part,
\begin{equation}
  S_\pm = R \pm S.
\end{equation}
The inverse transformation is given by
\begin{equation}
  R = \frac{1}{2} (S_+ + S_ -), \quad
  S = \frac{1}{2} (S_+ - S_ -).
  \label{eq:RSdefinition}
\end{equation}
When the mass is constant, the solutions $S_{\pm} = \mp \epsilon_{p} t + p x$
correspond to plane-wave solutions.
In that case, $R = p x$ and $S = -\epsilon_{p} t$.
Thus, the case $p = 0$ is singular in the sense that $R$ vanishes identically.
We will avoid this case as this leads to problems when using $R$ and $S$ as
coordinate transformations (see following subsection).

Using the above definition~\eqref{eq:RSdefinition} of $R$ and $S$ in the eikonal
equation and computing the sum and difference of the two equations, we find
\begin{equation}
  \begin{aligned}
    m^{2} &=
    {(\pd{t} R)}^{2} - {(\pd{x} R)}^{2} + {(\pd{t} S)}^{2} - {(\pd{x} S)}^{2}, \\
    0 &= (\pd{t} R) (\pd{t} S) - (\pd{x} R) (\pd{x} S).
  \end{aligned}
  \label{eq:eikonaleqsumdiff}
\end{equation}
We solve the latter equation for $\pd{t} R$ and obtain
\begin{equation}
  \pd{t} R = (\pd{x} R) \frac{\pd{x} S}{\pd{t} S}
  = \frac{\pd{x} R}{\pd{t} S} \pd{x} S
  = \lambda \pd{x} S,
\end{equation}
where we have introduced the abbreviation $\lambda$ which will be more
convenient later on.

Inserting this into to the first equation in~\eqref{eq:eikonaleqsumdiff} we get
\begin{equation}
  \lambda^{2} = 1 - \frac{m^{2}}{{(\pd{t} S)}^{2} - {(\pd{x} S)}^{2}}.
\end{equation}
The coefficients in the equations for $\alpha$ and $\beta$ finally are
\begin{equation}
  \begin{aligned}
    \Pi_\mu^\pm &= \pd{\mu} R \pm \pd{\mu} S, \\
    N_\pm &= \frac{1}{\sqrt{2m (m \mp \pd{t} R - \pd{t} S)}}, \\
    \kappa^\mu &= \frac{\sqrt{1 - \lambda^2}}{m} \varepsilon^{\mu \nu} \pd{\nu} S.
  \end{aligned}
\end{equation}

\subsection{Coordinate transformation}

Somewhat similar to the method of characteristics mentioned in the previous
section, we want to introduce new coordinates which simplify the evolution
equations~\eqref{eq:alphabetaexact} for the Bogoliubov coefficients.
The rapidly oscillating exponential contains the difference of the phases
$S=(S_+ - S_ -)/2$ and hence we choose one coordinate (the new time coordinate)
in this direction.
In order to have the same dimension as time, define the new time coordinate $s$
via $s = S(t, x)/m_0$ where $m_0 = \lim_{t \to -\infty} m$ is the asymptotic
value of the mass.
To simplify scalar products, the new spatial coordinate $r$ should be locally
orthogonal to $s$.
Inspecting the equations above, we find that this is automatically satisfied if
we define $r = R(t, x)/m_0$ in complete analogy.

Then we have
\begin{equation}
  \pd{s} S = m_{0}, \quad
  \pd{r} S = 0, \quad
  \pd{s} R = 0, \quad
  \pd{r} R = m_{0}
\end{equation}
and thus
\begin{equation}
  \Pi_{s}^{\pm} = \pm m_{0}, \quad
  \Pi_{r}^{\pm} = m_0.
\end{equation}
The components of the inverse metric tensor in $r$-$s$ coordinates are then
given by
\begin{equation}
  \begin{aligned}
    g^{ss} &= {(\pd{t} s)}^2 - {(\pd{x} s)}^2
    = \frac{1}{1 - \lambda^2} {\biggl( \frac{m}{m_0} \biggr)}^2, \\
    g^{rr} &= {(\pd{t} r)}^2 - {(\pd{x} r)}^2
    = -\frac{\lambda^2}{1 - \lambda^2} {\biggl( \frac{m}{m_0} \biggr)}^2, \\
    g^{rs} &= g^{sr} = (\pd{t} s) (\pd{t} r) - (\pd{x} s) (\pd{x} r) = 0,
  \end{aligned}
\end{equation}
where we see explicitly that the coordinates $r$ and $s$ are indeed locally
orthogonal.

Finally, the components of the Levi-Civita tensor are
\begin{equation}
  \varepsilon^{ss} = \varepsilon^{rr} = 0, \qquad
  \varepsilon^{sr} = -\varepsilon^{rs}
  = \frac{\lambda}{1 - \lambda^2} {\biggl( \frac{m}{m_0} \biggr)}^2.
\end{equation}
Additionally, we need to introduce the covariant derivative $\nabla_\mu v_\nu =
\pd{\mu} v_\nu - \Gamma^\lambda_{\mu \nu} v_\lambda$ where $\Gamma^\lambda_{\mu
  \nu}$ are the Christoffel symbols of the second kind.
The relevant derivatives that are needed in the
equations~\eqref{eq:alphabetaexact} for $\alpha$ and $\beta$ are
\begin{equation}
  \begin{aligned}
    \nabla_{\mu} \Pi_{s}^{\pm}
    = \underbrace{\pd{\mu} \Pi_{s}^{\pm}}_{\mathclap{= \pm \pd{\mu} m_0 = 0}}
    - \Gamma^\nu_{\mu s} \Pi_{\nu}^{\pm}
    = -m_{0} \bigl( \pm \Gamma^s_{\mu s} + \Gamma^r_{\mu s} \bigr), \\
    \nabla_{\mu} \Pi_{r}^{\pm}
    = \underbrace{\pd{\mu} \Pi_{r}^{\pm}}_{\mathclap{= \pd{\mu} m_0 = 0}}
    - \Gamma^\nu_{\mu r} \Pi_{\nu}^{\pm}
    = -m_{0} \bigl( \pm \Gamma^s_{\mu r} + \Gamma^r_{\mu r} \bigr).
  \end{aligned}
\end{equation}
Furthermore, we rescale $\alpha$ and $\beta$ according to
\begin{equation}
  \alpha = \tilde{\alpha} \sqrt{\lambda m}, \quad
  \beta = \tilde{\beta} \sqrt{\lambda m}.
  \label{eq:alphabetatilde}
\end{equation}
Again, we assume non-vanishing $p \neq 0$ as this would be singular for $p = 0$
because $\lambda \propto p$. Finally, after several manipulations and
simplifications, we get as equations for $\alpha$ and $\beta$
\begin{equation}
  \begin{aligned}
    &\pd{s} \tilde{\alpha} - \lambda^{2} \pd{r} \tilde{\alpha}
    - \lambda^{2} \tilde{\alpha} \pd{r} \ln \lambda \\
    &= -i e^{-2iS} \lambda \sqrt{1-\lambda^{2}} \biggl(
      \pd{r} \tilde{\beta} - \frac{1}{2} \tilde{\beta} \chi_{\beta} \biggr), \\[1ex]
    &\pd{s} \tilde{\beta} + \lambda^{2} \pd{r} \tilde{\beta}
    + \lambda^{2} \tilde{\beta} \pd{r} \ln \lambda \\
    &= i e^{2iS} \lambda \sqrt{1-\lambda^{2}} \biggl(
      \pd{r} \tilde{\alpha} - \frac{1}{2} \tilde{\alpha} \chi_{\alpha} \biggr)
  \end{aligned}%
  \label{eq:alphabetamassrs}
\end{equation}
with the abbreviations
\begin{equation}
  \begin{aligned}
    \chi_{\beta} &= \frac{1}{1-\lambda^{2}} \pd{s} \ln \lambda - 2 \pd{r} \ln m
    - \frac{1}{1-\lambda^{2}} \pd{r} \ln \lambda, \\
    \chi_{\alpha} &= \frac{1}{1-\lambda^{2}} \pd{s} \ln \lambda
      - \frac{1-2\lambda^{2}}{1-\lambda^{2}} \pd{r} \ln \lambda.
  \end{aligned}
\end{equation}
These equations~\eqref{eq:alphabetamassrs} are still exact, but they have
several advantages in comparison to the original Dirac
equation~\eqref{eq:diraceq}.
First, as in the purely time-dependent case, the rapidly oscillating phase
$e^{\pm 2iS}$ is a function of the new time-coordinate $s$ only.
Thus, they might also be advantageous for numerical simulations, especially when
the transformation from $(x,t)$ to $(r,s)$ coordinates can be implemented
efficiently.
Second, if $\lambda$ is small enough (see below) such that we may neglect terms
of order $\lambda^2$, these equations~\eqref{eq:alphabetamassrs} can be
approximated by
\begin{equation}
  \begin{aligned}
    \pd{s} \tilde{\alpha} &= -i e^{-2iS} \lambda  \biggl(
    \pd{r} \tilde{\beta} - \frac{1}{2} \tilde{\beta} \chi_{\beta} \biggr)
    + \mathcal{O} (\lambda^2), \\[1ex]
    \pd{s} \tilde{\beta} &= i e^{2iS} \lambda  \biggl(
    \pd{r} \tilde{\alpha} - \frac{1}{2} \tilde{\alpha} \chi_{\alpha} \biggr)
    + \mathcal{O} (\lambda^2).
  \end{aligned}%
  \label{eq:alphabetamassrs-approx}
\end{equation}
Third, the relevant case of $\alpha \gg \beta$, we see that $\tilde{\alpha}$
does approximately not evolve with time $s$, but stays nearly constant
$\tilde{\alpha} = \tilde{\alpha}(r)$, which fits to the picture of the
characteristics.
This suggests the picture of a wave packet $\alpha(r) e^{i m_{0} r + i m_{0} s}$
moving along curves of constant $r$ (i.e.\ in $s$-direction) whose shape is
given by $\alpha(r)$.
Going back to Cartesian coordinates $t$ and $x$ this corresponds to a wave
packet traveling at varying speed with the form of the wave packet changing over
time (i.e.\ becoming wider or narrower).
Then, we may solve the evolution equation for $\beta$ by integrating over $s$
for fixed values of $r$.
For each value of $r$, we have then the same situation as in the purely
time-dependent case, i.e., the pair creation exponent will be determined by the
complex value of $S$ at the first relevant singularity in the complex $s$-plane.

Note that this requires re-writing all functions of $t$ and $x$ as functions of
$s$ and $r$.
Then, for all fixed (real) values of $r$, one should analytically continue in
$s$ and find the singularities in the complex $s$-plane.
Since this procedure can only be applied fully analytically to special cases, we
develop a suitable approximation scheme based on weak spatial dependencies in
the following.

Another approach that could be considered is an inverse one (see
also~\cite{Oertel2015}):
If solutions $R$ and $S$ are given one can calculate the associated mass $m$
from the equations~\eqref{eq:eikonaleqsumdiff}.
These solutions could be obtained by choosing $S$ such that $R$ can be
calculated easily from~\eqref{eq:eikonaleqsumdiff}.

\section{Weakly space-dependent mass}%
\label{sec:weaklyspacedependent}

Consider a spacetime-dependent mass where the space-dependence is much weaker
than the time-dependence, i.e., $m = m(t, \varepsilon x) = m(t, \xi)$ with
$\varepsilon \ll 1$.
As before, we use the initial condition $S_\pm(t_{\mathrm{in}} \to -\infty, x) =
px = p \xi/\varepsilon$.
We then can expand the solutions of the eikonal
equation~\eqref{eq:eikonaleqsumdiff} in a power series for small $\varepsilon$,
\begin{equation}
  \begin{alignedat}{5}
    R &= \frac{1}{\varepsilon} R_{0} + &&R_{1}
    &&+ \varepsilon R_{2} &&+ \varepsilon^{2} R_{3} &&+ \dots, \\
    S &= &&S_{0} &&+ \varepsilon S_{1}
    &&+ \varepsilon^{2} S_{2} &&+ \dots
  \end{alignedat}
  \label{eq:RSexpansion}
\end{equation}
where $R_n$ and $S_n$, $n = 0, 1, 2, \dots$, are functions of $t$ and $\xi$.
Because only squares of the derivatives of $R$ and $S$ appear
in~\eqref{eq:eikonaleqsumdiff}, every second term in the expansions of $R$ and
$S$ vanishes, i.e.\ $R_{2n + 1} = S_{2n + 1} = 0$, $n = 0, 1, 2, \dots$.
To lowest order, we find
\begin{equation}
  R_0 = p \xi, \quad
  S_0 = - \int \diff{t} \sqrt{m^2 + p^2}.
  \label{eq:R0S0}
\end{equation}
These are exactly the same expressions as in the purely time-dependent case with
the only change that the mass $m$ now also depends on $x$ (or $\xi$).
The next non-vanishing terms are given by
\begin{equation}
\label{second-order-R-S}
  \begin{aligned}
    \pd{t} R_{2} &= \frac{(\pd{\xi} R_{0}) (\pd{\xi} S_{0})}{\pd{t} S_{0}}, \\
    \pd{t} S_{2} &= \frac{1}{2} \frac{{(\pd{\xi} S_{0})}^{2}}{\pd{t} S_{0}}
    + \frac{(\pd{\xi} R_{0})}{\pd{t} S_{0}} \Biggl[
      \pd{\xi} R_{2} - \frac{1}{2} \frac{\pd{\xi} R_{0}}{{(\pd{t} S_{0})}^{2}}
      \Biggr]
  \end{aligned}
\end{equation}
To simplify this further we assume that $p=\mathcal{O}(\varepsilon^2)$, i.e., $p
= \varepsilon^2 \tilde{p}$ where $\tilde{p} = \mathcal{O} (1)$.
Note that our WKB approximation is based on the assumption that the temporal
oscillations of $\exp\{iS_\pm\}$ are fast (of order $m$), the spatial variation
(and thus the momentum $p$) can be small.
In fact, pair creation is expected to be suppressed for large momenta $p$.
Inserting $p=\mathcal{O}(\varepsilon^2)$, we obtain
\begin{equation}
  \begin{aligned}
    R_2 &= \mathcal{O} (\varepsilon^2), \\
    S_2 &= \frac{1}{2} \int \diff{t} \frac{{(\pd{\xi} S_{0})}^{2}}{\pd{t} S_{0}}
    + \mathcal{O} (\varepsilon^2).
  \end{aligned}
\end{equation}
Using this approximation in~\eqref{eq:RSexpansion} we get
\begin{equation}
  \begin{aligned}
    R &= \varepsilon \tilde{p} \xi + \mathcal{O} (\varepsilon^{3}), \\
    S &= -\int \diff{t} \sqrt{m^2 + p^2}
    - \frac{1}{2} \varepsilon^{2} \int \diff{t} \frac{{(\pd{\xi} S_{0})}^{2}}{\sqrt{m^2 + p^2}}
    + \mathcal{O} (\varepsilon^{4}).
  \end{aligned}
\end{equation}
It should be noted that in the strict sense the square root $\sqrt{m^2 + p^2}$
should be expanded in a power series in $\varepsilon$ as well.
However, we assume that keeping this expression as it is will only enhance the
accuracy of our approximation.
Inserting these expansions into the definition of $\lambda$ we find
\begin{equation}
  \lambda = \frac{\pd{x} R}{\pd{t} S} = \varepsilon \frac{\pd{\xi} R}{\pd{t} S}
  = -\varepsilon^{2} \frac{\tilde{p}}{\sqrt{m^{2} + p^{2}}}
  + \mathcal{O} (\varepsilon^{4}).
\end{equation}
Hence, if we only keep terms up to order of $\varepsilon^{2}$
in~\eqref{eq:alphabetamassrs} the equations for $\alpha$ and $\beta$ are
\begin{equation}
  \begin{alignedat}{2}
    \pd{s} \tilde{\alpha} &= -&&i e^{-2iS} \lambda_{0} \biggl\{
      \pd{r} \tilde{\beta} - \frac{1}{2} \tilde{\beta} \bigl[
        \pd{s} \ln \lambda_{0} - \pd{r} \ln \bigl( \lambda_{0} m^{2} \bigr) \bigr]
      \biggr\}, \\
    \pd{s} \tilde{\beta} &= &&i e^{2iS} \lambda_{0} \biggl\{
      \pd{r} \tilde{\alpha} - \frac{1}{2} \tilde{\alpha} \bigl[
      \pd{s} \ln \lambda_{0} - \pd{r} \ln \lambda_{0} \bigr] \biggr\}
  \end{alignedat}
  \label{eq:alphabetarsweakly}
\end{equation}
where $\lambda_{0} = -p/\sqrt{m^2 + p^2}$ is the leading-order term of $\lambda$.

Again assuming the dominance of the positive frequency part $\alpha\gg\beta$
(i.e., that only few pairs are created), we find
$\tilde{\alpha}\approx\tilde{\alpha}(r)$.
Then $\tilde{\beta}_{\mathrm{out}}$ can be obtained from the second equation
in~\eqref{eq:alphabetarsweakly} by integrating over all $s$.
While performing that integral the other coordinate $r = p x/m_{0} + \mathcal{O}
(\varepsilon^{3})$ has to be held constant.
Fortunately, if we only keep terms up to order of $\varepsilon^{2}$, holding $r$
constant is the same as holding $x$ constant.

The integral is dominated by the singularity closest to the imaginary axis at
$s_{*} = S(t_{*}, x)$.
Typically, this will occur where $\lambda_{0}$ diverges, i.e.\ where
\begin{equation}
  m^{2}(t_{*}, x) + p^{2} = 0.
  \label{eq:dominantt}
\end{equation}
Here we assume that the function $m^{2}(t_{*},x)$ itself does not possess
singularities which are even closer to the real axis.
(This could be the case for dynamically assisted pair-creation, see,
e.g.,~\cite{Schutzhold2008, Fey2012, Linder2015}.)
In this case, they would determine $t_*$.

Thus, we expect $\beta_{\mathrm{out}}$ to behave like
\begin{equation}
  \beta_{\mathrm{out}}(x) \propto e^{2i S(t_{*}, x)}.
\end{equation}
The density of produced pairs will then be (see~\eqref{eq:positronnumberapprox})
\begin{equation}
  \abs{\beta_{\mathrm{out}}(x)}^{2} \propto e^{-2 \Im S(t_{*}, x)}.
\end{equation}
In complete analogy to the purely time-dependent case, we do not expect this
method to yield the correct prefactor due to the approximations made.

\subsection{Example: hyperbolic secant pulse}

As an example for an only weakly space-dependent mass we consider
\begin{equation}
  m(t, \xi) = m_{0} \sqrt{1 +
    {\biggl[ \frac{f(\omega t) g(\omega \xi)}{\gamma} \biggr]}^{2}}
  \label{eq:weaklyspacedependentmass}
\end{equation}
which is similar to~\eqref{eq:timedependentmass} but with an additional
space-dependent function $g(\chi)$.
In complete analogy to Eq.~\eqref{eq:timedependentmass}, we assume $\omega\ll
m_0$ and sufficiently small $\gamma$ in order to be in the WKB regime, the limit
of large $\gamma$ corresponds to the perturbative regime.

\begin{figure}
  \includegraphics{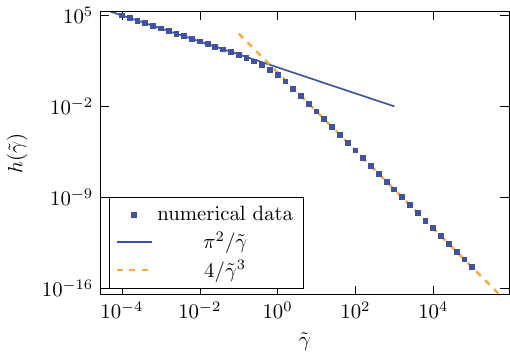}
  \caption{\label{fig:htildegamma}Log-log plot of numerically calculated
    function $h(\tilde{\gamma})$ as given in~\eqref{eq:htildegamma}. For small
    $\gamma \ll 1$, $h(\tilde{\gamma})$ approaches $\pi^{2}/\tilde{\gamma}$,
    whereas for large $\gamma \gg 1$ it behaves like $4/\tilde{\gamma}^{3}$.}
\end{figure}

We again use $f(\tau) = \sech \tau$.
Solutions to~\eqref{eq:dominantt} here are the same as in the time-dependent
case~\eqref{eq:taustar},
\begin{equation}
  \tau_{*} = \arcosh\biggl( \pm \frac{i}{\tilde{\gamma}} \biggr)
  = \ln \Biggl[ \frac{1}{\abs{\tilde{\gamma}}}
    + \sqrt{1 + \biggl( \frac{1}{\tilde{\gamma}^{2}} \biggr)} \Biggr]
  - i \frac{\pi}{2}
\end{equation}
with the only difference that now
\begin{equation}
  \tilde{\gamma} = \frac{\gamma}{g(\omega \xi)}
  \sqrt{1 + {\Bigl( \frac{p}{m_{0}} \Bigl)}^{2}}
\end{equation}
depends on $\xi$ (or, equivalently, $x$).
Comparing $S_{0}$ from~\eqref{eq:R0S0} with $\varphi_{p}$ from the
time-dependent case~\eqref{eq:varphipmass} we see that they are equal up to an
overall sign, i.e.\ $S_{0} = -\varphi_{p}$ and thus the lowest-order
contribution to the exponent of the number of produced pairs
\begin{equation}
  -4 \Im S_{0}(t_{*}, x) = -2 \pi \frac{m_{0}}{\omega}
  \sqrt{1 + {\Bigl( \frac{p}{m_{0}} \Bigl)}^{2}}
\end{equation}
is exactly the same as in the time-dependent case.
For the next-order contributions we have to calculate
\begin{equation}
  \begin{aligned}
    -4 \Im S_{2}(t_{*}, x) &= - 2 \Im \int_{-\infty}^{t_{*}} \diff{t}
    \frac{{(\pd{\xi} S_{0})}^{2}}{\pd{t} S_{0}} \\
    &= -\pi \frac{m_{0}}{\omega} \frac{{[ g'(\varepsilon \omega x) ]}^{2}}{\abs{g(\varepsilon \omega x)}}
    \frac{1}{\gamma} h(\tilde{\gamma})
    \label{eq:next-order}
  \end{aligned}
\end{equation}
with the dimensionless function $h$ depending on $\tilde{\gamma}$ only
\begin{widetext}
  \begin{equation}
    h(\tilde{\gamma}) = \Re \int_{0}^{1} \diff{u} \frac{{\left\{
        \arctan \left[ \dfrac{1}{\abs{\tilde{\gamma}}} \dfrac{%
            \cos \bigl( \pi u/2 \bigr) - i \sqrt{1 + \tilde{\gamma}^{2}}
            \sin \bigl( \pi u/2 \bigr)}{\sqrt{%
              (1 + \tilde{\gamma}^{2}/2) [ 1 + \cos ( \pi u ) ]
              - i \sqrt{1 + \tilde{\gamma}^{2}} \sin ( \pi u )}} \right]
        + \arctan \biggl( \dfrac{1}{\abs{\tilde{\gamma}}} \biggr) \right\}}^{2}}
    {\abs{\tilde{\gamma}} \sqrt{1 + \dfrac{1}{
    \cos ( \pi u ) + \tilde{\gamma}^{2}[ 1 + \cos ( \pi u ) ]/2
    - i \sqrt{1 + \tilde{\gamma}^{2}} \sin ( \pi u )}}}.
    \label{eq:htildegamma}
  \end{equation}
\end{widetext}
Note that because $\tilde{\gamma} = \tilde{\gamma}(\xi, p)$ this still depends
on the momentum $p$ and the spatial coordinate $\xi$.
This integral cannot be solved exactly in terms of elementary functions, but we
may obtain the asymptotics.
If we expand $h(\tilde{\gamma})$ in a series for small $\tilde{\gamma}$, we find
\begin{equation}
  h(\tilde{\gamma}) = \frac{1}{\tilde{\gamma}} \bigl[ \pi^{2}
    + \mathcal{O} ( \tilde{\gamma} ) \bigr].
\end{equation}
For large $\tilde{\gamma} \gg 1$, the integrand~\eqref{eq:htildegamma} decays
with $4/\tilde{\gamma}^{-3}$.
Note, however, that this limit corresponds to the perturbative regime, where the
WKB eventually breaks down.
To test this behavior, we calculated the function $h(\tilde{\gamma})$
numerically, see Fig.~\ref{fig:htildegamma}.

Because $h(\tilde{\gamma}) > 0$ for all values of $\tilde{\gamma}$, the
next-order contribution always decreases the pair-creation exponent, i.e., its
absolute value increases, thus reducing the number of produced pairs.
This is qualitatively consistent with the numerical results from~\cite{Gies2005}
using the worldline formalism.
There it was found that the locally constant field approximation overestimates
the true pair production probability, at least in the case of a Sauter
potential.

Consequently, we see that, to this order of approximation, the density of
produced pairs will be at its maximum where $g'(\omega \xi)$ vanishes.
Thus, both minima and maxima of the pulse may give significant contributions to
the number of produced pairs (compare~\cite{Gies2005}) as both are saddle points
of the spatial integral in~\eqref{eq:positronnumberapprox}.
However, the exact contribution depends on the prefactor in
$\beta_{\mathrm{out}}(x)$ which we have not calculated here.

Qualitatively, the momentum dependence of the total number $N(p)$ of produced
pairs will be the same as in the purely time-dependent case, i.e., quadratically
$N(p)\sim p^2$ for small $p$ and exponentially suppressed for large $p$.
The main effect of the spatial dependence of $m(t,x)$ will be an overall
reduction of the total amount of $N(p)$, due to the reduced pair-creation volume
or length and the correction~\eqref{eq:next-order} to the exponent.

\subsection{Higher momenta}

After Eq.~\eqref{second-order-R-S}, we used the low-momentum approximation
$p=\mathcal{O}(\varepsilon^2)$ in order to simplify the subsequent expressions.
This was sufficient for calculating the lowest-order
correction~\eqref{eq:next-order} to the pair-creation exponent which shows that
the spatial dependence tends to decrease the pair-creation probability.
However, as the mass varies on length scales on the order
$\mathcal{O}(1/\varepsilon)$, one might expect that further intersecting effects
occur on momentum scales of the order $p=\mathcal{O}(\varepsilon)$.
Thus, let us briefly discuss this case.
According to Eq.~\eqref{second-order-R-S}, $R_2$ can no longer be neglected
\begin{equation}
\pd{t} R_{2} = \frac{p}{\sqrt{m^2+p^2}}
\int \diff{t}\,\frac{\pd{\xi} m}{\sqrt{m^2 + p^2}}
\,,
\end{equation}
which implies that the coordinates $R$ and $x$ are no longer equivalent.
This complicates the analytical continuation because fixed and real values of
$R$ do not correspond to fixed and real values of $x$ (for complex $t$).

Furthermore, $\lambda$ is now less suppressed $\lambda=\mathcal{O}(\varepsilon)$
which implies that reaching the desired accuracy of
$\mathcal{O}(\varepsilon^2)$, one should keep the quadratic terms
$\mathcal{O}(\lambda^2)$ in the evolution equations~\eqref{eq:alphabetamassrs},
which adds further complications.
Of course, these more complicated equations can also be solved within a
consistent expansion in $\varepsilon$, but the resulting expressions will be
much more involved than those presented here.

For very large momenta $p$, on the other hand, one would expect that the results
simplify again because the locally homogeneous field approximation along the
particles worldline should become a good approximation.

\section{Conclusions \& Outlook}

Calculating the creation of particle pairs by truly spacetime-dependent external
fields (such as gravitational or electromagnetic fields) in the non-perturbative
regime is a very challenging task.
For purely time-dependent fields, a very powerful method to estimate the
pair-creation exponent is the WKB approximation.
In this work, we propose a generalization of this approach to truly
spacetime-dependent background fields, which is based on solutions of the
relativistic eikonal equation~\eqref{eq:eikonaleq}.
For fields that only depend on either time or a spatial coordinate, our method
reproduces the known results, see Sec.~\ref{sec:knowncases}.

One of the first obstacles we encounter is the problem of caustics.
They indicate that the eikonal equation~\eqref{eq:eikonaleq} in truly
spacetime-dependent background fields does not have globally differentiable
solutions in general, in contrast to the purely time-dependent case.
However, if the spatial dependence is sufficiently weak compared to the temporal
variation of the background, these caustics are well separated from the
space-time region of particle creation and thus do not spoil our approach, see
Sec.~\ref{sec:caustics}.

Then, via a transformation to adapted coordinates $r$ and $s$, the Dirac
equation in the presence of a spacetime-dependent mass $m(t,x)$ can be mapped
exactly to the equations~\eqref{eq:alphabetamassrs} for the Bogoliubov
coefficients.
These equations have several advantages and could also be suitable for improved
numerical simulation schemes.
In the low-momentum approximation $\lambda\ll1$, they simplify
to~\eqref{eq:alphabetamassrs-approx}.
Then, via the usual assumption that the positive frequency part dominates
$\alpha\gg\beta$, we may estimate the Bogoliubov coefficient $\beta$ associated
to pair creation via a simple integral over the new time coordinate $s$ in
complete analogy to the purely time-dependent case.
Thus, as in the purely time-dependent case, the pair-creation exponent is
determined by the first singularity in the complex $s$ plane.

Finally, consistent with our assumption to avoid caustics, we consider the case
that the spatial dependence is much weaker than the temporal variation and
employ an expansion in terms of the relative strength $\varepsilon$ of the
spatial dependence in Sec.~\ref{sec:weaklyspacedependent}.
To leading order, we obtain a result which is analogous to the locally constant
field approximation:
At each point $x$ in space, we simply have to integrate the evolution equation
for $\beta(t,x)$ over time -- in complete analogy to the purely time-dependent
case (as if we had a spatially homogeneous background).
In analogy to the locally constant field approximation, this leading order could
be referred to as locally homogeneous field approximation.

Calculating the next-to-leading order correction~\eqref{eq:next-order} to the
pair-creation exponent (for our example), we find that the spatial dependence
tends to decrease the pair-creation probability -- which is qualitatively
consistent with the behavior for the Sauter-Schwinger effect in a inhomogeneous
electric field, see, e.g.,~\cite{Dunne2005}.
Note that this next-to-leading order correction vanishes at maxima and minima of
the pulse, where $g'(\Omega\xi)$ is zero.

We expect that other field configurations where the dependence on one spacetime
coordinate is only weak can be treated similarly (e.g., tunneling through a
weakly time-dependent barrier or a light-front field pulse depending on $x_{+}$
plus a pulse only weakly dependent on $x_{-}$).
In the presence of an electromagnetic field $A_\mu$, one obtains formally the
same equations~\eqref{eq:alphabetaexact} for $\alpha$ and $\beta$, but the
subsequent steps such as the transformation to new coordinates $r$ and $s$ are
more involved.
It is still possible to use $S/m_{0}$ and $R/m_{0}$ as coordinates but they are
not locally orthogonal anymore.
Alternatively, one can obtain the coordinate $s = S/m_{0}$ in a similar way as
before and then construct another locally orthogonal coordinate but the
equations for the Bogoliubov coefficients $\alpha$ and $\beta$ become more
sophisticated nevertheless~\cite{OertelPhd}.
However, the main strategy should also be applicable in this case.

\acknowledgments

We thank anonymous referees for their constructive comments.
R.S.~acknowledges support by DFG (German Research Foundation), grant 278162697
(SFB 1242).

\appendix

\section{Pair production}%
\label{sec:pairproduction}

We want to give the relevant expressions for calculating the number of produced
pairs from solutions of~\eqref{eq:alphabetaexact}; see also textbooks
like~\cite{Fradkin1991, Greiner1985}.
Assuming that any field is switched off initially (i.e.\ for $t \to -\infty$)
and finally (i.e.\ for $t \to \infty$), the fermionic field operator
$\hat{\Psi}$ can be expanded in terms of one of two basis systems $\{
\psi_\mathrm{in}^\pm(p; t, x) \}$ or $\{ \psi_\mathrm{out}^\pm(p; t, x) \}$
\begin{equation}
  \begin{aligned}
    &\hat{\Psi} = \!\int \!\! \diff{p} \sqrt{\frac{m_{\mathrm{in}}}{2 \pi \epsilon_{p}^{\mathrm{in}}}}
    \bigl[
      \hat{a}_\mathrm{in}(p) \psi_\mathrm{in}^+(p; t, x)
      + \hat{b}_\mathrm{in}^\dagger(p) \psi_\mathrm{in}^-(p; t, x)
      \bigr] \\
    &\!\!= \!\! \int \!\! \diff{p} \! \sqrt{\frac{m_{\mathrm{out}}}{2 \pi \epsilon_{p}^{\mathrm{out}}}}
    \bigl[
      \hat{a}_\mathrm{out}(p) \psi_\mathrm{out}^+(p; t, x)
      \! + \! \hat{b}_\mathrm{out}^\dagger(p) \psi_\mathrm{out}^-(p; t, x)
      \bigr],
  \end{aligned}
  \label{eq:modeexpansion}
\end{equation}
where
\begin{equation}
  \begin{aligned}
    \epsilon_{p}^{\mathrm{in}}
    &= \sqrt{m_{\mathrm{in}}^{2} + {(p + qA_{1}^{\mathrm{in}})}^{2}}, \\
    \epsilon_{p}^{\mathrm{out}}
    &= \sqrt{m_{\mathrm{out}}^{2} + {(p + qA_{1}^{\mathrm{out}})}^{2}}
  \end{aligned}
\end{equation}
and $\hat{a}_\mathrm{in}(j)$, $\hat{b}_\mathrm{in}(j)$ and
$\hat{a}_\mathrm{out}(j)$, $\hat{b}_\mathrm{out}(j)$ are the initial and final
electron and positron annihilation operators, respectively.
The quantities $m_{\mathrm{in}}$ and $m_{\mathrm{out}}$ are the values of the
mass initially and finally, respectively, and similar for $A_{1}^{\mathrm{in}}$
and $A_{1}^{\mathrm{out}}$.
The functions $\{ \psi_\mathrm{in}^\pm(p; t, x) \}$ correspond to plane-wave
solutions for $t \to -\infty$ while the functions $\{ \psi_\mathrm{out}^\pm(p;
t, x) \}$ correspond to plane-wave solutions for $t \to \infty$; the
superscripts $+$ and $-$ denote positive and negative energy respectively.
These functions are complete,
\begin{equation}
  \begin{aligned}
    \sum_{\kappa} \int \diff{p} \frac{m_{\mathrm{in}}}{2 \pi
      \epsilon_{p}^{\mathrm{in}}} \psi_{\mathrm{in}}^{\kappa}(p; t, x) \,
        {(\psi_{\mathrm{in}}^{\kappa})}^{\dagger}(p; t, x')
        &= \delta(x - x'), \\
    \sum_{\kappa} \int \diff{p} \frac{m_{\mathrm{out}}}{2 \pi
      \epsilon_{p}^{\mathrm{out}}} \psi_{\mathrm{out}}^{\kappa}(p; t, x) \,
        {(\psi_{\mathrm{out}}^{\kappa})}^{\dagger}(p; t, x')
        &= \delta(x - x'),
  \end{aligned}
\end{equation}
and orthonormal,
\begin{equation}
  \begin{aligned}
    \bigl( \psi_{\mathrm{in}}^{\kappa}(p), \psi_{\mathrm{in}}^{\lambda}(p') \bigr)
    &= 2 \pi \frac{\epsilon_{p}^{\mathrm{in}}}{m_{\mathrm{in}}}
    \delta_{\kappa \lambda} \delta(p - p'), \\
    \bigl( \psi_{\mathrm{out}}^{\kappa}(p), \psi_{\mathrm{out}}^{\lambda}(p') \bigr)
    &= 2 \pi \frac{\epsilon_{p}^{\mathrm{out}}}{m_{\mathrm{out}}}
    \delta_{\kappa \lambda} \delta(p - p'),
  \end{aligned}
\end{equation}
where $( \cdot, \cdot )$ is the usual inner product defined as
\begin{equation}
  ( \phi, \psi ) = \int \! \diff{x} \: \phi^\dagger (t, x) \, \psi(t, x).
  \label{eq:innerproduct}
\end{equation}
Observe that
\begin{equation}
  \begin{aligned}
    \hat{a}_{\mathrm{in}}(p) &= \sqrt{\frac{m_{\mathrm{in}}}{2 \pi \epsilon_{p}^{\mathrm{in}}}}
    \bigl( \psi_{\mathrm{in}}^{+}(p), \hat{\Psi} \bigr), \\
    \hat{b}_{\mathrm{in}}^{\dagger}(p) &= \sqrt{\frac{m_{\mathrm{in}}}{2 \pi \epsilon_{p}^{\mathrm{in}}}}
    \bigl( \psi_{\mathrm{in}}^{-}(p), \hat{\Psi} \bigr), \\
    \hat{a}_{\mathrm{out}}(p) &= \sqrt{\frac{m_{\mathrm{out}}}{2 \pi \epsilon_{p}^{\mathrm{out}}}}
    \bigl( \psi_{\mathrm{out}}^{+}(p), \hat{\Psi} \bigr), \\
    \hat{b}_{\mathrm{out}}^{\dagger}(p) &= \sqrt{\frac{m_{\mathrm{out}}}{2 \pi \epsilon_{p}^{\mathrm{out}}}}
    \bigl( \psi_{\mathrm{out}}^{-}(p), \hat{\Psi} \bigr).
  \end{aligned}
\end{equation}
Then by using the respective other expansion of the field operator, one finds
the Bogoliubov transformation between the in- and out-operators
\begin{equation}
  \begin{aligned}
    \hat{a}_\mathrm{in}(p) &= \int \! \diff{p'} \Bigl[
      B_{p p'}^{+ +} \hat{a}_\mathrm{out}(p')
      + B_{p p'}^{+ -} \hat{b}_\mathrm{out}^\dagger(p')
      \Bigr], \\
    \hat{b}_\mathrm{in}^\dagger(p) &= \int \! \diff{p'} \Bigl[
      B_{p p'}^{- +} \hat{a}_\mathrm{out}(p')
      + B_{p p'}^{- -} \hat{b}_\mathrm{out}^\dagger(p')
      \Bigr], \\
    \hat{a}_\mathrm{out}(p) &= \int \! \diff{p'} \Bigl[
      \co{(B_{p' p}^{+ +})} \hat{a}_\mathrm{in}(p')
      + \co{(B_{p' p}^{- +})} \hat{b}_\mathrm{in}^\dagger(p')
      \Bigr], \\
    \hat{b}_\mathrm{out}^\dagger(p) &= \int \! \diff{p'} \Bigl[
      \co{(B_{p' p}^{+ -})} \hat{a}_\mathrm{in}(p')
      + \co{(B_{p' p}^{- -})} \hat{b}_\mathrm{in}^\dagger(p')
      \Bigr],
  \end{aligned}
\end{equation}
with the Bogoliubov coefficients
\begin{equation}
  B_{p p'}^{\kappa \lambda} =
  \frac{1}{2 \pi} \sqrt{\frac{m_{\mathrm{in}} m_{\mathrm{out}}}{\epsilon_{p}^{\mathrm{in}} \epsilon_{p'}^{\mathrm{out}}}}
  \bigl( \psi_\mathrm{in}^\kappa(p), \psi_\mathrm{out}^\lambda(p') \bigr)
\end{equation}
Thus, the number of produced positrons with momentum $p$ is
\begin{equation}
  \begin{aligned}
    N_{e^+}(p) &= \bra{0_\mathrm{in}}
    \hat{b}_\mathrm{out}^\dagger(p) \hat{b}_\mathrm{out}(p) \ket{0_\mathrm{in}}
    = \int \! \diff{p'} \abs{B_{p' p}^{+ -}}^2 \\
    &= \int \! \diff{p'}
    \frac{m_{\mathrm{in}} m_{\mathrm{out}}}{{(2 \pi)}^2 \epsilon_{p'}^{\mathrm{in}} \epsilon_{p}^{\mathrm{out}}}
    \abs{\bigl( \psi_\mathrm{in}^+(p'), \psi_\mathrm{out}^-(p) \bigr)}^2.
  \end{aligned}
  \label{eq:positronnumber}
\end{equation}
Let us assume that we calculated a solution to~\eqref{eq:alphabetaexact} with
the boundary conditions $\lim_{t \to -\infty} \alpha = 1$ and $\lim_{t \to
  -\infty} \beta = 0$, i.e.\ only positive energy initially.
Additionally, $\lim_{t \to -\infty} S_\pm = p x$.
Then we can actually use the wave function in~\eqref{eq:eikonalexpansion} as
$\psi_\mathrm{in}^+$.
Asymptotically, we thus find
\begin{equation}
  \begin{aligned}
    \psi_\mathrm{in}^+(p; t, x) \overset{t \to \infty}{\longrightarrow} \:
    &\alpha_\mathrm{out}(p; x) u_+^\mathrm{out}(p; x) e^{i S_+^\mathrm{out}(p; x)} \\
    + &\beta_\mathrm{out}(p; x) v_+^\mathrm{out}(p; x) e^{i S_-^\mathrm{out}(p; x)}
  \end{aligned}
\end{equation}
where the quantities designated with ``out'' are the values of their respective
time-dependent quantities at $t \to \infty$.
Similarly we have
\begin{equation}
  \psi_\mathrm{out}^-(p; t, x) \overset{t \to \infty}{\longrightarrow} \:
      \tilde{v}_+^\mathrm{out}(p) e^{i p x}.
\end{equation}
The spinor $\tilde{v}_+^\mathrm{out}(p)$ is obtained from a spinor $v_+$ at
$t \to \infty$ where solutions $\tilde{S}_\pm$ with the boundary condition
$\lim_{t \to \infty} \tilde{S}_\pm = p x$ have been used.

Because the inner product~\eqref{eq:innerproduct} is time-independent we may
evaluate the one in~\eqref{eq:positronnumber} at any time, e.g.\ for $t \to
\infty$ we find
\begin{equation}
  \begin{aligned}
    &\bigl( \psi_\mathrm{in}^+(p'), \psi_\mathrm{out}^-(p) \bigr) \\
    &\! = \! \int \!\! \diff{x} \Bigl\{ \co{\alpha_\mathrm{out}}(p'; x)
      {(u_+^\mathrm{out})}^\dagger(p'; x) \tilde{v}_+^\mathrm{out}(p) e^{-i[S_+^\mathrm{out}(p'; x) - p x]} \\
      &\quad + \co{\beta_\mathrm{out}}(p'; x)
      {(v_+^\mathrm{out})}^\dagger(p'; x) \tilde{v}_+^\mathrm{out}(p) e^{-i[S_-^\mathrm{out}(p'; x) - p x]}
      \Bigr\}.
  \end{aligned}
  \label{eq:psipluspsiminus}
\end{equation}
In the time-dependent case, $\pd{x} \alpha = \pd{x} \beta = 0$ and the canonical
momentum $p$ is conserved, i.e.\ $S_{\pm}^{\mathrm{out}}(p; x) = p x \mp
\varphi(p)$ where $\varphi(p)$ is independent of $x$.
Thus, $\tilde{v}_+^\mathrm{out} = v_+^\mathrm{out}$ is independent of $x$, too,
and using the identities
\begin{equation}
  {(u_+^\mathrm{out})}^\dagger(p) v_+^\mathrm{out}(p) = 0, \quad
  {(v_+^\mathrm{out})}^\dagger(p) v_+^\mathrm{out}(p) = \frac{\epsilon_{p}^{\mathrm{out}}}{m_{\mathrm{out}}},
\end{equation}
we find
\begin{equation}
  \bigl( \psi_\mathrm{in}^+(p'), \psi_\mathrm{out}^-(p) \bigr)
  = \frac{2 \pi \epsilon_{p}^{\mathrm{out}}}{m_{\mathrm{out}}} \co{\beta_\mathrm{out}}(p) \delta(p' - p)
\end{equation}
and therefore
\begin{equation}
  B_{p' p}^{+ -} =
  \sqrt{\frac{m_{\mathrm{in}}}{m_{\mathrm{out}}} \frac{\epsilon_{p}^{\mathrm{out}}}{\epsilon_{p}^{\mathrm{in}}}}
  \co{\beta_\mathrm{out}}(p) \delta(p' - p).
\end{equation}
Similarly we can calculate
\begin{equation}
  B_{p' p}^{+ +} =
  \sqrt{\frac{m_{\mathrm{in}}}{m_{\mathrm{out}}} \frac{\epsilon_{p}^{\mathrm{out}}}{\epsilon_{p}^{\mathrm{in}}}}
  \co{\alpha_\mathrm{out}}(p) \delta(p' - p).
\end{equation}
Thus, in the purely time-dependent case the coefficients $\alpha_{\mathrm{out}}$
and $\beta_{\mathrm{out}}$ essentially are the Bogoliubov coefficients and we
get for the number of produced pairs
\begin{equation}
  N_{e^+}(p) = \frac{m_{\mathrm{in}}}{m_{\mathrm{out}}}
  \frac{\epsilon_{p}^{\mathrm{out}}}{\epsilon_{p}^{\mathrm{in}}}
  \abs{\beta_\mathrm{out}(p)}^2 \delta(0)
\end{equation}
where the divergent factor $\delta(0)$ is due to the infinite extent of the
field.

In the spacetime-dependent case, the integral in~\eqref{eq:psipluspsiminus} is
far more difficult to solve, as most factors depend on $x$.
Still, for only weakly space-dependent fields as in
Section~\ref{sec:weaklyspacedependent} we assume that the dominant contribution
comes from a term similar to the one in the time-dependent case,
\begin{equation}
  N_{e^+}(p) \approx \frac{m_{\mathrm{in}}}{m_{\mathrm{out}}}
  \frac{\epsilon_{p}^{\mathrm{out}}}{\epsilon_{p}^{\mathrm{in}}}
  \int \diff{x} \abs{\beta_\mathrm{out}(p; x)}^2
  \label{eq:positronnumberapprox}
\end{equation}
which essentially is just the same expression as in the time-dependent case but
with $\delta(0)$ replaced with a spatial integral.
This is only a good approximation if scattering to other modes is low.

\section{Estimation of caustics}%
\label{sec:causticsestimation}

Using the method of characteristics, a first-order partial differential equation
may be turned into a set of first-order ordinary differential equations (see
e.g.~\cite{Evans1998} for a mathematical derivation of the method).
In our case, we can also use the following equivalent set of ordinary
differential equations
\begin{equation}
  \begin{alignedat}{2}
    \ddot{t}(\tau) &= &&\frac{2}{m_{0}^{2}} \pd{t} m^{2}, \\
    \ddot{x}(\tau) &= -&&\frac{2}{m_{0}^{2}} \pd{x} m^{2}, \\
    \dot{z}(\tau) &= &&\frac{2}{m_{0}} m^{2} = \frac{m_{0}}{2} \bigl( \dot{t}^{2} - \dot{x}^{2} \bigr),
  \end{alignedat}
\end{equation}
where $z(\tau) = S(t(\tau), x(\tau))$.
We use the boundary condition that at $t = t_{0}$ the solution $S$ is a plane
wave with positive energy which translates to the initial conditions
\begin{equation}
  \begin{aligned}
    t(\tau_{0}) &= t_{0}, \quad \dot{t}(\tau_{0}) = -2 \frac{\epsilon_{p}}{m_{0}}, \\
    x(\tau_{0}) &= x_{0}, \quad \dot{x}(\tau_{0}) = -2 \frac{p}{m_{0}}, \\
    z(\tau_{0}) &= -\omega t_{0} + p x_{0}.
  \end{aligned}
\end{equation}
These equations were solved numerically to obtain the characteristic curves in
fig.~\ref{fig:sechmasscharacteristics}.
Changing the parameter $\tau$ moves along a particular characteristic curve that
is specified by the starting position $x_{0}$.

We want to estimate the position of the intersection of two neighboring curves
analytically for $p = 0$.
For times prior to the pulse in the mass (where the mass is constant), the
(projected) characteristic curves are parallel to each other and their
parametrization is given by
\begin{equation}
  t = t_{0} - 2 \frac{\epsilon_{p}}{m_{0}} \tau, \quad
  x = x_{0} - 2 \frac{p}{m_{0}} \tau.
  \label{eq:initialcharacteristics}
\end{equation}
The curves are deflected when they reach the region of the pulse.
This deflection is manifest in a change of a curve's slope $\diff{x}/\diff{t}$
after passing the region of non-constant mass.
The above equations imply that the change of the slope with the parameter $\tau$
is
\begin{equation}
  \dd{\tau} \frac{\diff{x}}{\diff{t}}
  = \dd{\tau} \frac{\dot{x}}{\dot{t}}
  = -\frac{2}{\dot{t}} \frac{\pd{x} m^{2}}{m_{0}^{2}}
  - \frac{2 \dot{x}}{\dot{t}^{2}} \frac{\pd{t} m^{2}}{m_{0}^{2}}.
  \label{eq:characteristicsslope}
\end{equation}
To approximate the change in the slope we use the initial form of the
characteristic curves~\eqref{eq:initialcharacteristics}
in~\eqref{eq:characteristicsslope}.
We expect this to be a good approximation if $m$ is only weakly space-dependent,
i.e.\ the time scale on which the value of the mass changes is much smaller than
its length scale.
For $p = 0$ this approximation yields
\begin{equation}
  \dd{\tau} \frac{\diff{x}}{\diff{t}}
  \approx \frac{1}{m_{0}^{2}} \pd{x} m^{2}(t_{0} - 2 \tau, x_{0}).
\end{equation}
Thus, the slope after passing the region of non-constant mass is approximately
\begin{equation}
  \left. \frac{\diff{x}}{\diff{t}} \right|_{t \to \infty}
  \approx -\frac{1}{2 m_{0}^{2}} \int_{-\infty}^{\infty} \!\! \diff{t} \; \pd{x} m^{2}(t, x_{0}).
\end{equation}
Hence, a characteristic starting at $x = x_{0}$ will have the form
\begin{equation}
  x_{\mathrm{after}}(x_{0}; t) \approx x_{0}
  - \frac{t}{2 m_{0}^{2}} \int_{-\infty}^{\infty} \!\! \diff{t} \; \pd{x} m^{2}(t, x_{0})
\end{equation}
after passing the pulse.
The intersection of this characteristic and the one starting at $x_{0} + \delta$
is at
\begin{equation}
  t = \frac{2 m_{0}^{2} \delta}{%
    \int_{-\infty}^{\infty} \!\! \diff{t} \bigl[
    \pd{x} m^{2}(t, x_{0} + \delta) - \pd{x} m^{2}(t, x_{0}) \bigr]}.
\end{equation}
For $\delta \to 0$ this gives the intersection of two neighboring curves
\begin{equation}
  t = \frac{2 m_{0}^{2}}{%
    \int_{-\infty}^{\infty} \!\! \diff{t} \pd[2]{x} m^{2}(t, x_{0})}.
\end{equation}
Consequently, the focal point or onset of the caustic surface is where this is
at its minimum with respect to $x_{0}$.
For a weakly space-dependent mass of the
form~\eqref{eq:weaklyspacedependentmass} we get
\begin{equation}
  \begin{aligned}
    &\frac{1}{2 m_{0}^{2}} \int_{-\infty}^{\infty} \!\! \diff{t} \pd[2]{x} m^{2}(t, x_{0}) = \\
    &\frac{\varepsilon^{2} \omega^{2}}{\gamma^{2} \omega}
    \bigl\{ g(\varepsilon \omega x_{0}) g''(\varepsilon \omega x_{0}) + {\bigl[ g'(\varepsilon \omega x_{0}) \bigr]}^{2} \bigr\}
    \int_{-\infty}^{\infty} \diff{\tau} {\bigl[ f(\tau) \bigr]}^{2}
  \end{aligned}
\end{equation}
which immediately leads to the proportionality given in~\eqref{eq:caustictf}.
For $f(\tau) = \sech \tau$ and $g(\chi) = \sech \chi$ we find the minimum to be
\begin{equation}
  t_{f} = \frac{3}{2} \frac{\gamma^{2} \omega}{\varepsilon^{2} \omega^{2}}.
\end{equation}

\end{document}